\documentclass[apjl]{emulateapj}

\def\beq{\begin{equation}}
\def\eeq{\end{equation}}

\shorttitle {Magnetic Towers Driven by Magnetars}
\shortauthors{D.~A.~Uzdensky \& A.~I.~MacFadyen}

\begin{document}

\title{Magnetar-Driven Magnetic Tower as a Model for Gamma-Ray Bursts
and Asymmetric Supernovae}
\author{Dmitri A.\ Uzdensky\altaffilmark{1} \& 
Andrew I.\ MacFadyen\altaffilmark{2,3}}
\altaffiltext{1}{Princeton University, Department of Astrophysical Sciences,
Peyton Hall, Princeton, NJ 08544 --- Center for Magnetic Self-Organization
(CMSO); {\tt uzdensky@astro.princeton.edu}.}
\altaffiltext{2}{Institute for Advanced Study, Princeton, NJ 08540;
{\tt aim@ias.edu}.}
\altaffiltext{3}{Dept.\ of Physics, New York University, New York, NY 10003.}


\begin{abstract}
We consider a newly-born millisecond magnetar, focusing on its interaction
with the dense stellar plasma in which it is initially embedded. We argue
that the confining pressure and inertia of the surrounding plasma acts to
collimate the magnetar's Poynting-flux-dominated outflow into tightly beamed
jets and increases its magnetic luminosity. We propose this process as an
essential ingredient in the magnetar model for gamma-ray burst and asymmetric
supernova central engines. We introduce the ``pulsar-in-a-cavity'' as an
important model problem representing a magnetized rotating neutron star
inside a collapsing star. We describe its essential properties and derive
simple estimates for the evolution of the magnetic field and the resulting
spin-down power. We find that the infalling stellar mantle confines the
magnetosphere, enabling a gradual build-up of the toroidal magnetic field 
due to continuous twisting. The growing magnetic pressure eventually becomes
dominant, resulting in a magnetically-driven explosion. The initial phase of
the explosion is quasi-isotropic, potentially exposing a sufficient amount of
material to $^{56}$Ni-producing temperatures to result in a bright supernova.
However, if significant expansion of the star occurs prior to the explosion,
then very little~$^{56}$Ni is produced and no supernova is expected. 
In either case, hoop stress subsequently collimates the magnetically-dominated 
outflow, leading to the formation of a magnetic tower. After the star explodes,
the decrease in bounding pressure causes the magnetic outflow to become less 
beamed. However, episodes of late fallback can reform the beamed outflow, 
which may be responsible for late X-ray flares.
\end{abstract}    
\vspace{0.1 in}



\section{Introduction}
\label{sec-intro}


In the classical collapsar scenario for long-duration gamma-ray bursts (GRBs), 
the core of a rotating massive star collapses to form a black hole, whereas 
the overlying stellar material possesses enough angular momentum to form an 
accretion disk that persists for at least several seconds, long enough for 
its jet to breakout from the star (Woosley~1993; Paczynski~1998; MacFadyen
\& Woosley~1999). This accretion disk---black hole system then acts as a 
central engine for the GRB. The power for the explosion comes both from 
accretion energy, released via neutrinos and perhaps via a magnetic mechanism 
(e.g., the magnetic tower mechanism as proposed by Uzdensky \& MacFadyen 2006),
and from the black-hole rotational energy, released via the Blandford--Znajek 
(1977) mechanism.

In the present paper we investigate an alternative scenario, in which the 
central object formed as a result the of core collapse is not a black hole, 
but rather a rapidly-rotating (millisecond) magnetar with a large-scale 
poloidal magnetic field of the order of $10^{15}$~G. Such a strong magnetic 
field can be produced, for example, by a turbulent $\alpha-\Omega$ dynamo 
driven by convection in a proto-neutron star (PNS) subject to neutrino 
cooling (Duncan \& Thompson~1992; Thompson \& Duncan~1993). 
An alternative possibility is that the progenitor core of about $10^4$~km 
has a magnetic field of order $10^9$~G, similar to the field levels actually 
observed in some white dwarf of similar size. When such a highly-magnetized 
core collapses into a neutron star of 10~km radius, flux freezing leads to 
amplification of the magnetic field to $10^{15}$~G, as discussed in Uzdensky 
\& MacFadyen (2006). In addition, calculations by Akiyama~et~al. (2003) have 
shown that turbulent dynamo driven by the magneto-rotational instability (MRI) 
in the collapsing differentially-rotating core is capable of producing 
$10^{15}-10^{16}$~G fields within about 100~km from the center, on the 
timescale of just a few tens of milliseconds after the bounce (see also 
Ardeljan~et~al.\ 2005). 
Whatever the origin of the very strong magnetic field in the PNS is, in this 
paper we shall take it for granted. Our main goal will be to investigate the 
role such a strong field plays in the explosion dynamics.

The idea of using a millisecond magnetar as a central engine for 
gamma-ray bursts has been first proposed by Usov (1992) in the context 
of accretion-induced collapse of a highly-magnetic ($10^9$~G) white dwarf 
and, independently, by Duncan \& Thompson (1992). It has been further 
developed and applied to different explosion scenarios by several authors 
(e.g., Thompson~1994; Yi \& Blackman 1998; Nakamura~1998; Spruit~1999; 
Wheeler~et~al.\ 2000, 2002; Ruderman~et~al.~2000; Lyutikov \& Blandford 2002; 
Thompson~et~al.~2004; Metzger~et~al.\ 2007). The present paper is also 
devoted to investigating the millisecond-magnetar scenario, but viewed 
within the overall context of a collapsing star.

At the most basic level, the main idea is that a GRB (or a supernova) 
explosion is powered by the magnetic extraction of rotational energy 
of the newly-born rapidly-rotating magnetar. 
This magnetic luminosity operates alongside the much stronger neutrino 
cooling, which is the main avenue for releasing the gravitational binding 
energy of the young, still-contracting neutron star. However, most of 
the neutrinos escape to infinity without sharing their energy with the 
stellar-envelope gas (unless their spectrum is strongly modified by 
coronal processes, see Ramirez-Ruiz \& Socrates~2005). Magnetic fields,
on the other hand, couple to the gas tightly and this makes them a very 
efficient explosion agent. Energetically, magnetic GRB models are usually 
quite plausible.
For example, assuming a typical surface magnetic field $B_*=10^{15}$~G, 
a rotation rate $\Omega_*=10^4\ {\rm sec}^{-1}$, and a radius $R_*=10$~km,
it is easy to see that the resulting basic energetics and timescales fall
just in the right ballpark to make the millisecond magnetar a plausible 
candidate for a GRB central engine (e.g., Thompson~1994). 
Indeed, the total rotational energy of a millisecond-period neutron star 
is $E_{\rm rot}\simeq 5 \cdot 10^{52}\ {\rm erg}$, which is more than enough 
to drive a long-duration GRB.
The time scale for the energy extraction can be estimated by dividing
this available energy by the total magnetic luminosity (the spin-down 
power), $L_{\rm magn}$. The latter can be roughly estimated by the usual 
pulsar luminosity formula $L_{\rm magn}\sim B_*^2\, R_*^6\, \Omega_*^4\, 
c^{-3}$, which for the above parameters yields $\sim 3\cdot 10^{50}\ {\rm 
erg\ sec}^{-1}$, corresponding to the characteristic timescale of order 
$100$~sec. Thus, from the point of view of the overall energetics and 
timescales, the millisecond-magnetar central engine is just a scaled-up 
version of the Ostriker \& Gunn (1971) model for regular pulsar-powered 
supernovae (with the magnetic field scaled up by three orders of magnitude 
and the time scaled down by six orders of magnitude).

It has to be noted, however, that the plausible overall energetics
and timescales are, by themselves, not sufficient for making a good
GRB central engine model. This is because there are some extra physical
requirements mandated by observations.
In particular, to make a successful GRB, the central engine has to be 
capable of producing an energetic outflow that is (1) ultra-relativistic; 
(2) highly-collimated; and (3) baryon-free. The pioneering works cited 
above have focused  mostly on the energetics and timescales, but not on 
the mechanisms for producing an outflow that satisfies these requirements 
(see, however, Wheeler~et~al.\ 2000 and Bucciantini~et~al.~2006 for a 
discussion of collimation). 
Also, most of these previous models, with the notable exceptions of
Wheeler~et~al.\ (2000) and Arons (2003), see~\S~\ref{subsec-const-pressure}, 
have considered a magnetar in isolation; that is, they have completely 
ignored the effect of any surrounding stellar gas on shaping the outflow. 
This may be a good approximation for the accretion-induced collapse 
of a white dwarf, but it is not appropriate in the collapsing-star 
scenario.

In contrast, in this paper we stress that the infalling stellar gas is 
still present during the explosion and needs to be taken into account.
Thus, an important new element that distinguishes our model from those 
previous works is the consideration of the interaction between a 
newly-born magnetar and the stellar plasma in which it is initially 
embedded. Specifically, we argue that the pressure and inertia (i.e., 
the ram pressure) of the surrounding stellar gas acts as a natural 
collimator forcing the magnetized outflow into two tightly beamed 
jets. It also plays a crucial role in magnetic extraction of 
rotational energy from the magnetar. 

In order to illustrate these ideas we introduce the ``Pulsar-in-a-Cavity'' 
problem as a basic-physics paradigm for this scenario. We describe this 
problem in detail in~\S~\ref{sec-pulsar-cavity}. We first give a general 
description of the problem and its various versions. Then, in \S~\ref
{subsec-fixed-cavity}, we consider the simplest special case of a rotating
force-free magnetosphere inside a fixed rigid cavity. In that section, we 
first demonstrate that differential rotation of the magnetic field lines 
is inevitably established inside the cavity, even if the pulsar itself is 
rotating uniformly; as a result, a strong toroidal magnetic field gradually 
builds up. We then study the long-term evolution of the magnetic field 
inside the cavity and show that the magnetic luminosity increases with 
time. We also show that a massive, non-force-free plasma strip unavoidably 
arises in the equatorial plane beyond the light cylinder. 
In \S~\ref{subsec-collimation} we discuss the subtle issue of hoop-stress 
collimation and argue that external confinement and differential rotation 
are two important ingredients for collimating relativistic Poynting-flux
dominated outflows. We then consider, in \S~\ref{subsec-const-pressure}, 
the case of a magnetosphere surrounded by a cavity with a fixed external 
pressure (instead of a fixed radius).

In~\S~\ref{sec-stalled-shock} we discuss a specific example relevant to 
the core collapse of a massive star: a cavity formed behind the stalled 
bounce-shock at the center of the collapsing star. The radius of the shock 
stays roughly stationary on the timescale for magnetic fields in the cavity 
to grow. At the same time, both the ram pressure of the gas falling onto 
the cavity and the neutrino energy deposition inside it decrease with time.
We therefore argue that at some point, a fraction of a second after bounce, 
the magnetic field will inevitably start to dominate the force balance, 
leading to a magnetically-driven explosion.

In~\S~\ref{sec-discussion}, we further explore some of the 
astrophysically-interesting aspects of our model. Thus, in
\S~\ref{subsec-stability}, we discuss the possibility of MHD
instabilities (e.g., kink) developing in the twisted magnetosphere
and their implications for our model. In \S~\ref{subsec-nickel} we 
address an important issue of $^{56}$Ni production and argue that 
the two-phase nature of the explosion in our model is well-suited 
to explain a large amount of $^{56}$Ni inferred from observations.
In~\S~\ref{subsec-restarting} we briefly discuss the possibility 
of restarting the GRB engine by the fall-back of the post-explosion
material. In~\S~\ref{subsec-preshaping} we describe an extension
to our model: a ``magnetar-in-a-tube'', motivated by the fact that 
the material along the rotation axis does not experience a centrifugal 
barrier and hence falls onto the PNS faster. In section~\S~\ref
{subsec-pulsar-kicks} we discuss the implications of our model for 
pulsar kicks. Finally, in~\S~\ref{subsec-numerical}, we suggest 
some directions for future numerical simulations of this problem. 
We draw our conclusions in~\S~\ref{sec-conclusions}.


\section{The Pulsar-in-a-Cavity Problem}
\label{sec-pulsar-cavity}

\begin{figure}
\plotone{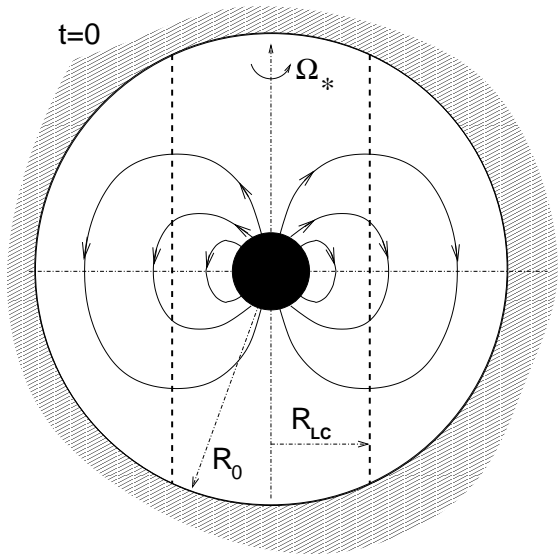}
\figcaption{Aligned pulsar inside an infinitely-conducting spherical 
cavity of radius~$R_0$ at $t=0$. The vertical dashed lines represent 
the pulsar's light cylinder of radius $R_{\rm LC}<R_0$. 
\label{fig-magnetar-1}}
\end{figure}

In order to understand how a millisecond magnetar central engine 
operates in the collapsar context, it is first necessary to consider 
the following basic physics problem: what happens when an axisymmetric 
pulsar is placed inside a conducting cavity filled with a low-density 
infinitely-conducting plasma (see~Fig.~\ref{fig-magnetar-1})? Specifically, 
we are interested in a situation where the cavity radius~$R_0$ is much larger 
than the pulsar light-cylinder radius~$R_{\rm LC}$. We call this idealized 
problem the {\it Pulsar-in-a-Cavity problem} (Uzdensky \& MacFadyen 2006)
and we propose it as the first essential step in building up a physical 
understanding of the problem.
It is a modification of the famous problem of an axisymmetric 
rotating magnetic dipole in free space, considered by Goldreich 
\& Julian (1969) as a model for an isolated pulsar's magnetosphere. 
In some aspects, it is also similar to the system considered by
Kardashev (1970), Ostriker (1970; unpublished), and by Ostriker 
\& Gunn (1971). 

We would like to point out that adding a cavity makes the problem, 
in a sense, less fundamental since the behavior of the system depends, 
in general, on the assumed physical properties of the cavity. To make 
the situation less arbitrary, we shall fix the electromagnetic properties 
of the cavity by assuming that its walls are perfectly conducting, as is 
the plasma that fills the cavity. We shall also assume that all the field 
lines close back to the pulsar inside the cavity, i.e., that there are no 
field lines connecting the pulsar to the cavity wall. 
This choice is most natural when the bulk of the magnetic flux has been 
produced by a dynamo operating inside the neutron star and then emerged 
through its surface (as is in the case of magnetars), as opposed to a 
situation where the pulsar field lines connect directly to the wall 
(i.e., to the outer stellar envelope, as considered, e.g., by Kardashev~1970 
and by Goldreich~et~al. 1971).
At the same time, we are still left with a lot of freedom regarding the 
mechanical properties of the cavity. Thus, we are dealing not with one 
unique problem, but instead with a whole class of problems. Correspondingly, 
we propose that the overall problem be treated as a sequence of test problems 
with increasingly more sophisticated treatment of the cavity boundary. 
Depending on the physical situation, this sequence may also represent 
various stages in the time evolution of the system.

For example, one can first consider the case where the cavity 
walls are rigid and have a fixed (e.g., spherical) shape (see 
\S~\ref{subsec-fixed-cavity}). 
This may represent the early stages of the system's evolution. 
Next, one can assume that the shape and the size of the cavity 
are not fixed but instead are governed by the pressure balance between 
the electromagnetic stress inside the cavity and a constant external
pressure outside (see \S~\ref{subsec-const-pressure}). This makes 
the set-up similar to Lynden-Bell's (1996) magnetic tower model. 
Thirdly, one can consider a pulsar in a fully dynamic environment 
of a collapsing star. The set-up of the latter problem is essentially
similar to that considered by Ostriker \& Gunn (1971).

All three versions of our pulsar-in-a-cavity problem are 
basic physics problems that ought to be solved if we are 
ever to understand how a millisecond magnetar works inside 
a collapsing star. Our understanding of these problems will 
benefit from rigorous mathematical analysis, but ultimately 
will most likely be achieved with the help of numerical 
simulations that are now becoming feasible. Whereas the 
first two problems represent perfect targets for relativistic 
force-free simulations, the third problem will most likely
require a full relativistic MHD simulation.

To set the stage for future numerical studies, and to be able to 
interpret their results, it is useful to get some basic qualitative 
understanding of the problem. Therefore, in this section we will sketch 
what we think is a plausible physical picture of the system's evolution 
and how it relates to our magnetic tower model for~GRBs (Uzdensky \& 
MacFadyen 2006).

In order to gain a more complete understanding of the interaction 
between the central magnetar and the surrounding stellar material, 
a full magnetohydrodynamic (MHD) description that includes plasma 
pressure and inertial effects will eventually be required. Of particular 
interest would be the confinement of the expanding magnetosphere by the 
surrounding plasma and the dynamical response of the star to the expanding 
magnetosphere at its center. The full-MHD approach is especially relevant 
if there is a strong wind driven off the~PNS by neutrinos and/or by the 
magneto-centrifugal mechanism, as considered by Thompson~et~al. (2004) 
and by Bucciantini~et~al.\ (2006). A useful simplification may come from 
noting that the main difference between the dense-plasma case and the 
relativistic force-free case is just the difference between the Alfv\'en 
speed and the speed of light (J.~Ostriker, private communication). Then, 
the MHD case may be treated similarly to the relativistic force-free case, 
but with the light cylinder replaced by a smaller Alfv\'en surface.

For simplicity, however, in this paper we shall restrict ourselves to 
the force-free case. That is, we shall assume that the plasma density 
inside the cavity is so low that electromagnetic forces dominate the 
dynamics almost everywhere inside the cavity (but outside the neutron 
star of course). The only exception is the part of the equatorial plane 
outside the light cylinder, where plasma inertia needs to be taken into 
account (see below).
While not realistic, given the large plasma densities present in the center
of a massive star, the force-free description may nonetheless reflect some
essential features of the full solution. It is of relevance especially for 
late phases of the evolution when the magnetic field outside the neutron 
star has been amplified to large values.

As we have already mentioned, we shall also assume that the plasma inside
the cavity can be accurately represented by an infinitely conducting fluid. 
We expect this key assumption to be well justified throughout most of the 
cavity, owing to the very large plasma densities and temperatures. Indeed, 
the high plasma density ensures that the plasma (including photons) is highly 
collisional and hence is well described by resistive MHD; this means that the 
resistivity due to particle-particle or photon-particle collisions dominates 
over all other non-ideal terms in generalized Ohm's law. On the other hand, 
because of the very high plasma temperature, the resistivity is actually 
quite small, i.e., the magnetic Reynolds number is very high. All this makes 
ideal MHD a good approximation in the environment of a collapsing star 
(see Uzdensky \& MacFadyen 2006 for more discussion). 
At the same time, we do acknowledge that this assumption may break down in 
some special regions, in particular, inside the equatorial plasma strip 
(see below) and at the cavity boundary, where various fluid instabilities 
may lead to enhanced turbulent energy dissipation. In any case, however, 
we expect ideal MHD to be much better justified inside a collapsing star 
than the force-free assumption. For this reason, in this paper we shall 
ignore any finite-resistivity effects, leaving them for a future study.
In addition, because of the very high density, the plasma inside 
the cavity is completely optically thick to electromagnetic radiation
and so the photons are tightly coupled to the gas. What this means is 
that there is no radiative (apart from possible neutrino cooling which 
we ignore in this basic problem) cooling in our system. Therefore, all 
the energy that is extracted from the pulsar stays inside the cavity, 
as long as we don't allow the cavity to expand.

Throughout most of this discussion we shall ignore all numerical factors, 
e.g., $4\pi$, etc. Also, we shall assume that the magnetar rotation 
rate~$\Omega_*$ stays approximately constant on the timescales under 
consideration. However, ultimately one will have to consider the effect 
of decreasing rotation rate as the pulsar slows down.

Finally, we would like to stress that there are important differences between 
the problem of an isolated pulsar magnetosphere and our pulsar-in-a-cavity 
problem. In particular, in the isolated pulsar case one usually seeks a 
steady state (although perhaps employing time-dependent simulations to 
achieve it). In the pulsar-in-a-cavity case, on the other hand, we don't 
expect a stationary solution; the problem is intrinsically time-dependent 
and it is the time evolution of the system that is of particular interest.
In addition, it is believed that the wind of a normal isolated pulsar crosses 
the fast magnetosonic surface somewhere far beyond the light cylinder and then 
reaches the termination shock. This is important because it implies that
the inner pulsar magnetosphere is causally disconnected from the outside;
in particular, the inner magnetosphere's structure and the pulsar spin-down
power cannot be influenced by the boundary conditions at very large distances.
In sharp contrast, our case of a magnetosphere enclosed within a finite-size
cavity is qualitatively different, because we now lack that huge separation of
radial scales. In practical terms, this means that we draw a dividing line
between the isolated pulsar magnetosphere and the bounded pulsar magnetosphere
based on the presence or absence of the fast magnetosonc surface inside the 
cavity. In particular, in our present study we are interested in the case 
of a cavity formed inside the stalled supernova shock (see~\S~\ref
{sec-stalled-shock}). Its radius may be about 100--200~km, i.e., 
only moderately larger than the light cylinder radius of a millisecond 
magnetar (about 30~km). Then, the plasma outflow just may not have enough 
range to reach the fast magnetosonic surface. As a result, our bounded pulsar 
magnetosphere always remains in causal contact with the outer boundary. 
Correspondingly, the inner structure of the magnetosphere and the pulsar 
spin-down power is affected by the confining cavity.


\subsection{Pulsar in a Fixed Spherical Cavity}
\label{subsec-fixed-cavity}

We start with our problem I, in which the walls of the cavity
are fixed. For definiteness, we take the cavity to be spherical 
in shape. The main results obtained in this section should also 
be approximately valid for the case of expanding (or contracting) 
cavity, as long as the expansion (contraction) speed is slow 
compared with the speed of light.

Let us try to think physically about how the magnetic field will 
evolve after the pulsar is spun-up instantaneously at $t=0$.
In the Goldreich--Julian (1969) model for an isolated pulsar, 
the field lines extending beyond the light cylinder bend backwards 
and tend to become open. [Strictly speaking, the field lines actually 
always close, but very far away, in the so-called ``boundary zone'' 
(Goldreich \& Julian 1969).] As long as the cavity boundary (the 
outer edge of the magnetosphere) lies outside the fast magnetosonic 
surface of the outflow, there is no feedback of this boundary on the 
inner magnetosphere. Then, the pulsar continuously spins down, losing 
its rotational energy and angular momentum to magnetic braking by 
these effectively-open field lines. In our case, on the other hand, 
such an immediate field-line opening is not possible since the entire 
magnetosphere is contained inside the cavity of a finite size. This is 
one of the most critical differences between the isolated pulsar case 
and our case.


\subsubsection{Development of Differential Rotation}
\label{subsubsec-differential-rotation}

One important point that one needs to take into account is the 
establishment of {\it differential rotation} in the magnetosphere. 
This is nontrivial, since, by assumption, the magnetar rotates 
uniformly. However, as we will now show, the field lines that 
extend beyond the pulsar light cylinder nevertheless necessarily 
undergo differential rotation. As a result, these field lines are 
continuously twisted and hence toroidal flux is continuously injected 
into the cavity. 

\begin{figure}
\plotone{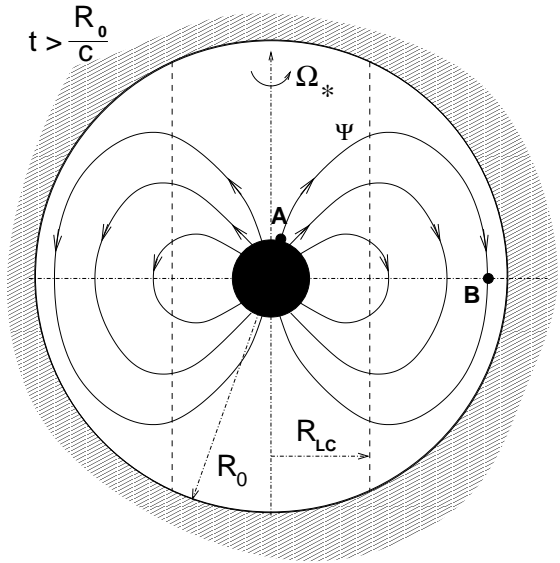}
\figcaption{Aligned pulsar inside an infinitely-conducting 
spherical cavity of radius~$R_0$. After a time of order the
light-crossing time~$R_0/c$, the poloidal field lines outside 
the light cylinder expand somewhat but still remain confined within 
the cavity. Because the toroidal magnetic field has to vanish 
at the equatorial midplane due to reflection symmetry, the field 
lines there cannot corotate with the star, $\Omega_B<c/R_B<\Omega_*$.
As a result, differential rotation is established in both hemispheres,
$\Delta\Omega=\Omega_*-\Omega_B\simeq\Omega_*$ (for $R_B\gg R_{\rm LC}$),
which leads to continuous generation of toroidal magnetic flux.
\label{fig-magnetar-2}}
\end{figure}

To see how this comes about, let us consider a field line~$\Psi$ 
(Fig.~\ref{fig-magnetar-2}) and compare the angular velocities at 
two points on this line: point~$A$, where the field line attaches 
to the pulsar, and point~$B$, where it intersects the equator.
Since this field line extends beyond the light cylinder, it cannot 
remain purely poloidal: toroidal field has to develop so that the 
plasma particles could slide backwards and out along the line, like
beads on a wire.  
This toroidal field leads to a continuous braking of the star 
so that there is an outward  flux of angular momentum and a 
Poynting flux of energy along the line. However, the toroidal 
magnetic field at point~$B$ has to be exactly zero because of 
the assumed reflection symmetry with respect to the midplane. 
Therefore, the plasma can no longer slide toroidally; this 
means that the toroidal velocity of the field line is equal 
to that of the plasma at this point. The angular momentum 
and rotational energy of the pulsar extracted by the magnetic 
field are partly accumulated and stored in the magnetic form 
and partly transferred to the equatorial plasma.
Thus, the material at point~$B$ is continuously torqued by the 
magnetic field. Then, since the confining wall prevents the
material from moving out freely in the radial direction, 
the toroidal velocity of the plasma becomes closer and closer 
to the speed of light. However, it can never exceed the speed 
of light; therefore, the plasma, and hence field-line, angular 
velocity at point~$B$ is bounded:
$\Omega_B \simeq c/R_B = \Omega_* R_{\rm LC}/R_B$. On the other 
hand, the angular velocity at point~$A$ is of course just the 
rotation rate of the pulsar: $\Omega_A = \Omega_*$.
This means that the field line experiences differential rotation 
at a rate $\Delta\Omega = \Omega_A-\Omega_B \geq \Omega_* (1-R_{\rm LC}/R_B)$.
For field lines that cross the equator well outside the light cylinder,
$R_B\gg R_{\rm LC}$, we then have $\Delta\Omega\approx \Omega_*$. 
Thus, differential rotation is established over a time-scale of 
order the light-crossing time across the cavity, $t_0\equiv R_0/c
\gg\Omega_*^{-1}$.

This differential rotation is important because it leads to a 
continuous toroidal stretching of the field lines and thus to 
a continuous injection of toroidal magnetic flux (of opposite 
signs) into the upper and lower hemispheres. Since all this 
toroidal flux has to be contained within a cavity of fixed 
size, the toroidal magnetic field at any given point grows, 
roughly speaking, linearly in time. This is in contrast with 
the pulsar in a free space, where, within a sphere of any given 
radius, a steady state is established on the time scale of order 
the light travel time across this radius. One worry that one might 
have in our case is the possibility of the kink instability as the 
magnetic field in the cavity becomes highly wound up. We address 
this issue in more detail in~\S~\ref{subsec-stability}.

Finally, the toroidal field reverses sharply across the equator, so there 
is a non-force-free equatorial current sheet that carries the radial return 
current back to (or from) the neutron star (see below).


\subsubsection{Magnetic field structure at late times}
\label{subsubsec-late-times}

Now let us try to estimate the toroidal field evolution and distribution
inside the cavity on long time scales ($t\equiv N t_0$, where $N \gg 1$, 
and $t_0\equiv R_0/c$ is the light crossing time across the cavity)
and at distances much larger than the light cylinder radius.

As we discussed above, because of the differential rotation, the bounded 
magnetosphere cannot be stationary: toroidal magnetic flux is constantly 
being injected into a finite volume. Hence, the toroidal field strength 
continuously increases, whereas the poloidal magnetic field does not. 
The poloidal electric field, $E_{\rm pol}$,  may become much larger 
than~$B_{\rm pol}$ but in any case cannot exceed the value~$B_{\rm pol} 
\Omega_* R_0 /c=B_{\rm pol} R_0/R_{\rm LC}$.
Thus, after several light-crossing times the magnetosphere outside 
the light cylinder becomes toroidal-field dominated: $B_\phi\gg 
E_{\rm pol}, B_{\rm pol}$. 

Next, even though the configuration is time-dependent, after many 
light-crossing times the evolution slows down. Indeed, the poloidal 
field structure readjusts (e.g., in response to a change in the 
toroidal field strength) on a time scale of order the fast-magnetosonic 
crossing time across the cavity; for a force-free plasma this coincides 
with~$t_0= R_0/c$. 
Since the toroidal flux grows linearly in time, the relative change 
in the toroidal field strength over~$\Delta t\sim t_0$ becomes small 
(of~order~$N^{-1}$) at late times, $t=N\, t_0$, $N\gg 1$. An approximate 
force-free equilibrium is then established separately in each of the two 
hemispheres, with poloidal current being approximately constant on poloidal 
flux surfaces: $I\simeq I(\Psi)$. The magnetic field structure in such an 
equilibrium is governed by the relativistic force-free Grad--Shafranov 
equation (aka the pulsar equation). 
In the limit where the toroidal magnetic field totally dominates the dynamics, 
this equation reduces to $II'(\Psi)=0$, so that the poloidal current function 
becomes independent of~$\Psi$: $I(\Psi)=I_0={\rm const}$. This corresponds 
to the vacuum field produced by a singular line current $I_0$ (which grows 
linearly in time) flowing along the rotation axis. The toroidal magnetic 
field is $B_\phi(t,R,Z)=I_0(t)/R$, i.e., $B_\phi={\rm constant}$ on 
cylinders, and the equilibrium can be described as the balance between the 
toroidal field tension and pressure. In other words, the ${\bf j\times B}$ 
force becomes relatively small inside the cavity, because the poloidal current 
becomes spatially separated from the toroidal magnetic field: it flows out of 
the pulsar along the axis (in both hemispheres), then as a surface current 
along the cavity walls, and finally returns to the pulsar along the 
non-force-free equatorial current sheet. The bulk of the magnetosphere 
is thus almost current-free. In this regard, the electric-current structure 
of the cavity is similar to that of the magnetic bubble considered by Lyutikov 
\& Blandford (2002, 2003) in their model for Poynting-flux dominated GRB 
outflows (although we apply our model deep inside the collapsing star,
that is, on different spatial and temporal scales compared with their 
model).

Let us now estimate the magnitude of the poloidal line current~$I_0(t)$
and hence the characteristic strength of the toroidal field in the cavity. 
We shall express magnetic quantities characterizing the field in the cavity 
in terms of the total poloidal magnetic flux that extends beyond the light 
cylinder, which we shall call~$\Psi_0$. Up to a factor of order unity, this 
flux can be estimated from the pure dipole magnetic field, i.e.,
\beq
\Psi_0 \sim \Psi_{\rm dipole}(R_{\rm LC}) = 
B_*\ {{R_*^3}\over{R_{\rm LC}}} \, .
\label{eq-Psi0}
\eeq 
This estimate is justified because inside the light cylinder the
poloidal field remains close to dipole. Moreover, even in the extreme
case of an unbounded, isolated pulsar magnetosphere, in which the field 
is completely open outside the light cylinder, the poloidal flux crossing
the light cylinder differs from the dipole formula only by a small amount 
(e.g., Contopoulos, Kazanas \& Fendt 1999; Komissarov~2006; McKinney~2006b; 
Spitkovsky~2006). Thus, this estimate should be quite good in our case as
well.

Now, what is the characteristic poloidal magnetic field strength 
in the cavity at distances $r\sim R_0$ ? Here the dipole formula 
($B_{\rm pol}\sim r^{-3}$), describing a fully-closed non-rotating 
field, and the split-monopole formula, describing the fully-open 
magnetosphere of an isolated pulsar, differ. In our case, all the
field lines are closed, i.e., they intersect the equator within~$R_0$,
so one might think that the dipole-field estimate should be more applicable.
However, as we show below, almost all of the field lines crossing the 
light cylinder actually intersect the equator in a narrow strip near 
the outer wall; therefore, the characteristic poloidal field at distances 
of order~$R_0$ from the center and off the equatorial plane should be 
estimated as
\beq
B_{\rm pol} \sim B_0 \equiv {\Psi_0\over{{R_0}^2}} \, .
\label{eq-B0-def}
\eeq
For $\Psi_0$ given by equation~(\ref{eq-Psi0}), this estimate gives 
a value $B_{\rm pol} \sim B_*\, (R_*^3/R_0^2 R_{\rm LC})$, which is 
by a factor $R_0/R_{\rm LC}$ larger than a pure dipole field at these 
distances.

Now let us estimate the poloidal current and the toroidal magnetic field.
In general, the poloidal current flowing through a region enclosed by an 
axisymmetric flux surface~$\Psi$ can be calculated by following the shape 
of a field line corresponding to~$\Psi$:
\beq
I(\Psi,t) = \Delta\Omega t \ \biggl[ \int\limits_\Psi 
{{dl_{\rm pol}}\over{B_{\rm pol}R^2(l_{\rm pol})}} \biggr]^{-1} \, ,
\label{eq-I-twist}
\eeq
where $l_{\rm pol}$ is the path-length along the poloidal field.
The main contribution to the integral comes from large distances,
$R\sim R_0$ and thus the integral can be estimated as being of
order~$R_0/\Psi_0$. Then, since $\Delta\Omega\simeq\Omega_*=c/R_{\rm LC}$,
the axial poloidal current can be estimated as
\beq
I_0(t) \sim \Omega_* t \ {\Psi_0\over{R_0}} \simeq
{\Psi_0\over{R_{\rm LC}}} \, {t\over t_0} \, .
\eeq
Thus we see that for $t\gg t_0$ the poloidal current becomes much 
stronger than the typical poloidal current in the unbounded pulsar 
magnetosphere ($I\sim \Psi_0/R_{\rm LC}$). Using the estimate~(\ref
{eq-Psi0}) for~$\Psi_0$, we can express $I_0$ as
\beq
I_0(t) \sim B_* \, {{R_*^3}\over{R_{\rm LC}^2}} \, {t\over t_0} \, .
\label{eq-I_0}
\eeq

Correspondingly, the characteristic toroidal magnetic field at distances 
of order~$R_0$ is 
\beq
B_\phi(R_0) = {I_0\over{R_0}} \simeq B_0 \Omega_* t \, ,
\label{eq-B_phi}
\eeq 
which is similar to the estimate presented by Kardashev (1970) 
for the toroidal field of a pulsar inside an expanding supernova
cavity.
We see that, after many light-crossing times across the cavity, 
$B_\phi(R_0)$ becomes much larger than the toroidal field of an 
isolated pulsar at these distances 
[$B_\phi^{\rm isolated} \sim \Psi_0/(R_0 R_{\rm LC}) =
B_0 (R_0/R_{\rm LC}) = B_0 \Omega_* t_0 \ll B_0 \Omega_* t $].

Finally, we would like to remark on how to determine the structure 
of the poloidal field, $\Psi(r,\theta)$. Usually, when studying 
steady-state axisymmetric magnetospheres, one uses an iterative 
procedure (e.g., Contopoulos~et~al.~1999). First, one makes a guess
for the poloidal current~$I(\Psi)$, plugs it into the Grad--Shafranov 
equation, and solves this equation for~$\Psi(r,\theta)$. Then one uses 
equation~(\ref{eq-I-twist}) to determine the new function~$I(\Psi)$ and 
repeats the steps until the procedure converges. In our case, however, 
this approach does not appear to be feasible, since to lowest order 
the Grad--Shafranov equation simply gives $I(\Psi)=I_0={\rm const}$. 
We therefore advocate for an inverted approach where the poloidal flux 
function is determined from equation~(\ref{eq-I-twist}). 
How to realize such an approach in practice is not clear.
One thing to note though is that this calculation should 
depend on $\Psi(R,Z=0)$ as a boundary condition, and this 
has to be determined from considering the redistribution 
of the poloidal flux across the equatorial midplane. This 
issue is discussed in the next subsection.


\subsubsection{Centrifugal Force in the Equatorial Plane}
\label{subsubsec-centrifugal}

\begin{figure}
\plotone{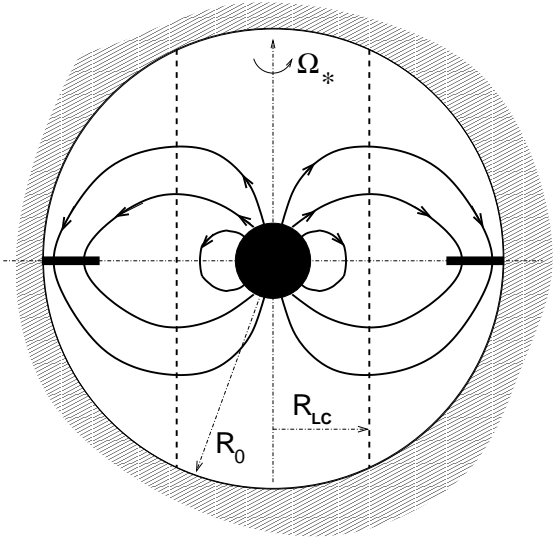}
\figcaption{At late times, the poloidal magnetic field is pressed against 
the wall by the centrifugal force of the rotating massive equatorial sheet. 
\label{fig-strip}}
\end{figure}

As we noted above, the magnetosphere outside the pulsar light cylinder 
cannot be entirely force-free. Because the toroidal magnetic field 
reverses across the equator (due to the assumed reflection symmetry), 
the magnetic field tension continuously accelerates the equatorial 
plasma in the toroidal direction. Correspondingly, this tension force 
performs mechanical work on the equatorial plasma and so a certain part 
of the rotational energy extracted from the pulsar by the magnetic field 
is deposited in the equatorial plane (the rest is stored in the bulk of 
the cavity as the toroidal magnetic field energy). 
Since the plasma in the equatorial plane rotates ultra-relativistically, 
the added energy leads to an increase in the relativistic ``mass'' of 
the plasma, $\Delta m\sim t^2$. An important consequence is that this 
relativistically-rotating massive equatorial sheet experiences an outward 
radial centrifugal force, $F_{\rm cent}$. 
This force cannot be balanced by the toroidal magnetic field 
because the latter is zero at the equator. Consequently, the 
equatorial plasma moves towards the wall and compresses poloidal 
magnetic field, until finally the centrifugal force is balanced 
by the ${\bf j\times B}$ force due to the non-force-free part of
the toroidal current $j_\phi$.%
\footnote
{The contribution from the the force-free part, $j_\phi^{\rm ff}(z=0)=
\rho_e v_\phi(z=0)$ is exactly canceled by the radial electric force, 
$\rho_e E_r$, provided that the ideal-MHD condition ${c \bf E=v\times B}$ 
holds in the equatorial strip.} 
Thus, the poloidal magnetic flux in the equatorial plane outside 
the light cylinder is pushed against the wall and is strongly 
concentrated in a narrow band of ever-decreasing width $d(t)\ll R_0$ 
near the wall (see Fig.~\ref{fig-strip}).
Because of this effect, we can expect nearly all the poloidal flux~$\Psi_0$ 
that extends beyond the light cylinder to cross the equator inside this strip, 
i.e., at cylindrical radii~$R\simeq R_0$.
At the same time, in the magnetosphere above and below the equatorial plane,
the poloidal field lines that emanate from this band have to fan out because 
they have to fill the cavity volume. Thus, the characteristic poloidal 
magnetic field in the cavity is of the order~$B_0=\Psi_0/R_0^2$ (see 
eqn.~\ref{eq-B0-def}) and is much weaker (by a factor of $d/R_0$) than 
in the equatorial strip.

Let us assess the centrifugal force quantitatively.
The total torque exerted on the massive equatorial strip by the magnetic 
field is given by $\tau(t)=\int I(\Psi,t) d\Psi \simeq 
I_0(t)\Psi_0$. 
Since the toroidal velocity of this strip is close to the speed of light,
the total work per unit time due to this torque (i.e., the total Poynting
flux that arrives at the strip) is $P_{\rm strip}\simeq \tau c/R_0=B_{\phi} 
c\Psi_0$.
This power goes into accelerating the rotation of plasma in the strip,
and some part of it may in principle dissipated into heat. 
Since the rotation here is already ultra-relativistic, the result 
of this acceleration is an increase in the rotation and/or thermal 
$\gamma$-factors, i.e., of the relativistic mass~$m$ of the plasma 
in the strip: $d/dt(mc^2)=P_{\rm strip}$. As a result, the relativistic 
mass grows with time as
\beq
m(t)c^2\sim\Omega_*^2 t^2 {{R_{\rm LC}}\over{R_0}}\,{\Psi_0^2\over{R_0}}\sim 
\biggl({t\over{t_0}}\biggr)^2\, {{R_0}\over{R_{\rm LC}}}\,{\Psi_0^2\over{R_0}}
\sim {{R_{\rm LC}}\over{R_0}}\, B_\phi^2(t) R_0^3 \, ,
\label{eq-strip-mass}
\eeq
that is, the plasma energy in the equatorial strip always remains 
small compared with the energy  $B_\phi^2(t) R_0^3$ stored in the 
toroidal magnetic field at these distances. The centrifugal force 
acting on the equatorial strip can be estimated as 
\beq
F_{\rm cent}(t) = {m(t)c^2\over{R_0}} \sim
B_0^2 R_0^2 \Omega_*^2 t^2\, {{R_{\rm LC}}\over{R_0}} \sim
B_\phi^2(t) R_0^2\, \biggl({{R_{\rm LC}}\over{R_0}}\biggr) \, .
\eeq
We see that this force grows quadratically with time, just 
as the toroidal field pressure, but always remains small 
(by a factor of $R_{\rm LC}/R_0\ll 1$) compared with the 
overall horizontal force exerted on the side wall by the 
toroidal field.

A detailed analysis of the internal structure of the massive equatorial
plasma strip, including its vertical structure, lies beyond the scope of 
this paper. However, we present here a simple estimate for the Lorentz 
factor due to rotation, $\gamma_{\rm rot}$, in terms of the strip width~$d$ 
and half-thickness~$h$. This estimate is derived under a certain very 
restrictive set of assumptions and serves for illustration only.

Let us consider the vertical force balance inside the strip in the
co-rotating frame, and let us neglect the contribution from electric 
force for simplicity. Then the toroidal magnetic field pressure outside 
the strip has to be balanced by the plasma pressure inside: 
$p_{\rm co} = B_\phi^2/8\pi$. 
Next, let us make the assumption that the plasma in the strip is a light 
relativistic fluid with the adiabatic index~4/3. Then, the co-moving energy 
density is $\rho_{\rm co}c^2 = 3 p_{\rm co}$. On the other hand, the total 
plasma energy $mc^2$ inside an annular strip of radius~$R_0$, width~$d$, 
and thickness~$2h$ can be written in the lab frame as 
$m c^2 = 4\pi R_0 dh\, \rho_{\rm co} c^2 \gamma_{\rm rot}^2$.  
By combining all these expressions with the equation~(\ref{eq-strip-mass}) 
for~$mc^2$, we find 
\beq \gamma_{\rm rot}^2 \sim {2\over 3}\, {{R_0 R_{\rm LC}}\over{dh}} \, .
\label{eq-gamma-rel-fluid}
\eeq

On the other hand, it may be possible that a significant amount of plasma 
accumulates in the equatorial strip or that the plasma there is strongly 
compressed by the toroidal field pressure. Then, the baryon number density 
$n_b$ may become so large that the co-moving energy density is dominated by 
the non-relativistic component, i.e., by the baryon rest-mass, 
$\rho_{\rm co}c^2 \simeq n_{b,\rm co} m_p c^2 \gg p_{\rm co}=B_\phi^2/8\pi$.
Since $n_{b,\rm co}=\gamma_{\rm rot}^{-1}\, n_b$, the condition that this is 
true can be written as
\beq
\gamma_{\rm rot}^{-1} \gg \sigma \equiv 
{{B_\phi^2}\over{4\pi n_b m_p c^2}} \, .
\label{eq-baryon-dominant-cond-1}
\eeq
Provided that we are in this regime, the total plasma energy in the strip 
is dominated by the kinetic energy of the baryons: $m c^2 = 4\pi R_0 dh\, 
\gamma_{\rm rot} n_b m_p c^2$. Then, using equation~(\ref{eq-strip-mass}), 
we get
\beq
\gamma_{\rm rot} \sim {{R_0 R_{\rm LC}}\over{dh}} \, \sigma_{\rm strip}\, ,
\label{eq-gamma-baryons}
\eeq
By substituting this expression into the condition~(\ref
{eq-baryon-dominant-cond-1}), we see that the co-moving 
energy density is dominated by the rest-mass of the baryons 
only when
\beq
\sigma \ll \sqrt{dh\over{R_{\rm LC} R_0}} \ll 1 \, .
\label{eq-baryon-dominant-cond-2}
\eeq
Correspondingly, we have 
\beq
\gamma_{\rm rot} \ll \sqrt{{R_0 R_{\rm LC}}\over{dh}} \, .
\eeq


\subsubsection{Magnetic Spin-down Power of a Pulsar in a Fixed Cavity}
\label{subsubsec-magn-power}

Another extremely important point is that the rate at which 
the magnetic field in a bounded magnetosphere extracts rotational 
energy from the central rotating conductor actually grows with time. 
This is because the magnetic torque per unit area is proportional to 
the toroidal field at the conductor's surface and the latter grows 
linearly with time. Thus, the magnetic power generated by a spinning 
pulsar inside a cavity increases linearly with time as long as the 
cavity does not expand (or expands slowly) and the spin rate of the 
pulsar stays constant. We can estimate the spin-down power as
\beq
P(t) = I(t) \Psi_0 \Omega_*  = \Omega_*^2 t\ {\Psi_0^2\over{R_0}} \sim
P_{\rm isolated}\, {ct\over{R_0}} \, ,
\eeq
where $P_{\rm isolated}\sim B_*^2 R_*^6\Omega_*^4/c^3$ is the spin-down power 
of an isolated, unbounded pulsar. As we see, after many light-crossing times,
the power of a pulsar-in-a-cavity greatly exceeds that of a classical isolated 
pulsar. This is our answer to the apparent paradox raised by Lyutikov (2006).

We thus emphasize that the energy extraction from a magnetar-in-a-cavity 
can be a run-away process. This is because the twisting of a magnetic field 
confined by an external boundary results in an increase in the field's 
strength at the light cylinder and hence in a growing  rate of energy 
extraction from the magnetar.

This effect can be attributed to a positive feedback that exists 
between the energy that has been already extracted from the pulsar, 
and the strength of the agent that extracts the energy (the toroidal
magnetic field). Namely, most of the extracted energy is stored
in the toroidal magnetic field, and since the volume occupied by 
this field is kept finite, the toroidal field strength increases
with time. Because the magnetosphere remains in a quasi-equilibrium, 
the toroidal field constantly readjusts everywhere, including within 
the light cylinder. In other words, because the system is not hyperbolic 
but elliptic, the inner magnetosphere feels the presence of the outer 
confining wall. In particular, the toroidal field at the very surface 
of the pulsar increases linearly with time, and hence so does the spin-down 
torque exerted by the magnetic field on the pulsar. This picture is similar 
to what is happening in the combustion chamber of a rocket, for example.
In that case, the gas temperature and pressure increase as the chemical
energy of the fuel is released in the combustion process. At the same
time, the rate at which fuel burning occurs increases with an increase
in the ambient temperature. As a result, rapid and efficient burning 
demands high pressure and hence a strong confining chamber capable of 
withstanding this pressure. Similarly, in our case of a pulsar placed
inside a cavity, the presence of strong cavity walls leads to an 
increased energy extraction rate from the pulsar.

In a realistic situation, this steady power growth might not last 
indefinitely. It may be limited, for example, by the development of 
the kink instability, which would result in the conversion of the 
toroidal flux to poloidal flux and to partial dissipation of magnetic 
energy (see \S~\ref{subsec-stability} for more discussion).


\subsection{Hoop-stress collimation: contrast with the isolated pulsar}
\label{subsec-collimation}

The toroidal field generated by the differential rotation exerts 
a constantly-growing pressure on the cavity walls. If we now relax 
the assumption that the walls are fixed and allow them to move, this 
pressure will make the cavity inflate. We then want to understand how 
rapidly such inflation will proceed and whether it will be isotropic 
or, say, collimated along the axis. We discuss the collimation issue 
in this subsection.

Generally speaking, since the toroidal field pressure in the lateral 
direction is partly negated by the field's tension (the hoop stress), 
which has no vertical component, one may expect the resulting expansion 
to be predominantly vertical. However, notice that here we are interested
in a situation where the (differential) rotation is relativistic: 
$\Delta\Omega R_0\sim \Omega_* R_0\gg c$. 
On the other hand, Lynden-Bell's (1996) magnetic tower model, for example, 
was developed for the non-relativistic regime. It is well-known that 
hoop-stress collimation is not a trivial issue in the relativistic case. 
Thus, it is not immediately obvious that the hoop-stress collimation 
mechanism can be applied to the pulsar-in-a-cavity scenario considered 
in this paper. We therefore would like to discuss this issue in some 
detail here. 

At first, one might think that there should be no problem
collimating the outflow: the magnetic field is predominantly
toroidal even without differential rotation. And it is the 
toroidal field's hoop stress that is usually credited for 
collimating astrophysical jets. However, as is well known, 
hoop-stress collimation does not work as well when applied 
to ultra-relativistic magnetically-dominated outflows, as 
it does in the non-relativistic case.
The quintessential example of this lack of collimation 
is the isolated-pulsar wind inside the termination shock.
The basic reason for this difficulty is the decollimating force 
due to the poloidal electric field, $E_{\rm pol}$. Indeed, in the 
case of an {\it unbounded} relativistic uniformly-rotating force-free 
magnetosphere (e.g., an isolated aligned pulsar magnetosphere) 
in a steady state, the poloidal electric and toroidal magnetic 
fields have to be nearly equal in strength at large distances 
from the central axis (Goldreich \& Julian 1969). 
Importantly, it turns out that this balance can be realized in 
an uncollimated, quasi-spherical poloidal magnetic field configuration; 
an excellent example of this is Michel's (1973) split-monopole solution.
A rough argument explaining this lack of hoop-stress collimation in the
relativistic-rotation case goes as follows. 
Let us consider an uncollimated field configuration; the poloidal magnetic 
field is open outside the light cylinder and has a split-monopole geometry, 
i.e., drops off with distance as~$r^{-2}$. 
In a steady state, the poloidal electric field is $E_{\rm pol}=
B_{\rm pol}\, R/R_{\rm LC}$ where~$R$ is the cylindrical radius. 
It therefore drops off along radial rays as~$r^{-1}$. But the 
toroidal magnetic field also drops off as~$r^{-1}$. Moreover, 
at the light cylinder~$E_{\rm pol}$ and~$B_\phi$ are comparable:
$E_{\rm pol}=B_{\rm pol}\sim B_\phi$. Since outside the light 
cylinder they both decrease as the same power of~$r$, they remain 
comparable to each other (both being much larger than~$B_{\rm pol}$)
at large distances. Moreover, as Goldreich \& Julian (1969) showed, 
$E_{\rm pol}$ and~$B_\phi$ even become equal asymptotically as 
$r\rightarrow\infty$. The bottom line is that a quasi-spherical 
relativistic force-free equilibrium can be established as a balance 
between the collimating pinch force (the sum of the toroidal magnetic 
field pressure and its tension) and the opposing electric force. 
Hoop-stress collimation is suppressed as a result of this balance.

\begin{figure}
\plotone{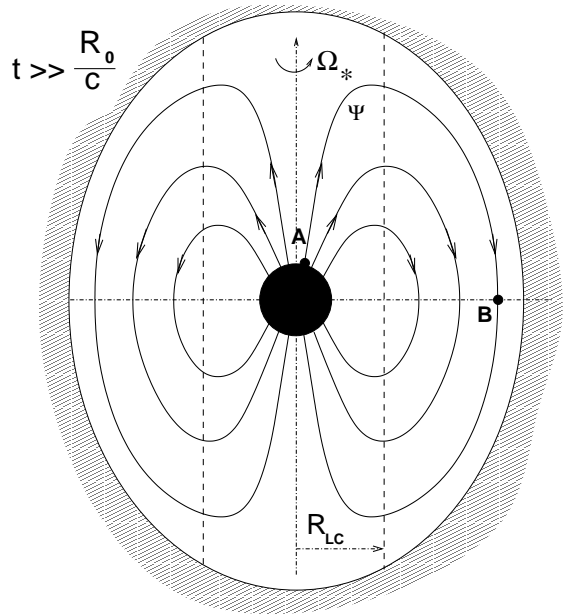}
\figcaption{Axisymmetric pulsar inside a cavity.
After many light-crossing times, the magnetosphere becomes 
toroidal-field dominated. Because of hoop-stress, the magnetic 
stress on the cavity becomes strongly concentrated near the axis.
This leads to a predominantly vertical, collimated expansion; 
a magnetic tower forms.
\label{fig-magnetar-3}}
\end{figure}

Now, in the case of a rotating magnetosphere enclosed inside a rigid
cavity of a fixed radius~$R_0>R_{\rm LC}$, the situation is different
and hoop-stress collimation can in fact work.  Indeed, as we showed
above, after many light-crossing times ($t\gg R_0/c$), the toroidal
magnetic field filling the cavity becomes stronger than both~$B_{\rm
pol}$ and~$E_{\rm pol}$, in contrast to the isolated pulsar case. 
Moreover, this toroidal field is distributed nonuniformly; it is
basically inversely proportional to the cylindrical radius. 
Correspondingly, the stress exerted by the toroidal magnetic field on the 
cavity walls is also nonuniform: the magnetic pressure pushing vertically 
against the top and bottom walls is much higher that the lateral magnetic 
pressure acting on the side walls. Therefore, if we now allow the cavity 
to expand under this pressure, we expect any subsequent expansion to be 
mostly vertical (see Fig.~\ref{fig-magnetar-3}), at least as long as the 
expansion velocity is slow compared with the speed of light. Then we 
effectively find ourselves in a situation similar to the non-relativistic 
magnetic tower proposed by Lynden-Bell (1996). We therefore envision that 
the eventual, long-term result of this process will be the creation of a 
pair of oppositely-directed magnetic towers (Uzdensky \& MacFadyen 2006). 
The interaction of the expanding towers with the surrounding stellar 
envelope aids in their confinement, similarly to jet collimation seen 
in hydrodynamical simulations of the collapsar model (MacFadyen \& 
Woosley 1999; Aloy~et~al.~2001; MacFadyen, Woosley \& Heger 2001; 
Zhang, Woosley \& MacFadyen 2003). In the scenario considered in the 
present paper, these towers are driven not by a differentially-rotating 
disk, but by a rapidly-rotating magnetar. This suggests that considering 
the magnetosphere of a pulsar inside a cylindrical, as opposed to spherical, 
cavity may represent yet another interesting and important problem for future 
study (see~\S~\ref{subsec-preshaping}).

An important element in the above discussion is the fact that 
the electric field is small compared with the toroidal magnetic 
field. This is directly related to the fact that the toroidal 
field is generated not as a part of an outgoing large-scale 
electromagnetic wave driven by the pulsar rotation, but as a 
result of differential rotation. This observation points to the 
important role played by {\it differential} rotation (as opposed 
to uniform relativistic rotation) in collimating relativistic 
force-free outflows.

\subsection{Pulsar magnetosphere confined by a constant external pressure}
\label{subsec-const-pressure}

Let us now consider the case when the pulsar magnetosphere is confined
by some fixed and uniform external gas pressure, $P_{\rm ext}$, instead 
of a cavity of fixed radius~$R_0$. We are interested in this particular 
set-up because it is closest to that considered by Lynden-Bell in his 
original magnetic tower paper (Lynden-Bell~1996), and we here want to 
compare his non-relativistic disk model with a pulsar in a similar setting.

Like Lynden-Bell, let us assume that the external pressure is 
weak compared with the magnetic field pressure $B_*^2/8\pi$ in 
the immediate vicinity of the rotating conductor. Moreover, 
because we are interested in exploring relativistic effects, 
we want our pulsar magnetosphere to be able to expand well 
beyond the light cylinder. 
Therefore, we shall also assume that the external pressure is small 
compared with the magnetic pressure of a pure dipole field at the light 
cylinder: $8\pi P_{\rm ext}\ll B_{\rm dipole}^2(R_{\rm LC})\sim B_*^2\, 
(R_*/R_{\rm LC})^6 \ll B_*^2$.

Let us imagine, as is frequently done in time-dependent pulsar
magnetosphere studies (e.g., Komissarov~2006; McKinney~2006b; 
Spitkovsky~2006), that we start with a non-rotating star with 
a dipole field and then spin it up suddenly at $t=0$.
The initial evolution of the magnetic field is then similar 
to that of an isolated pulsar: the field lines that extend 
beyond the light cylinder start to wind up and expand radially 
at the speed of light, i.e., $R_0\simeq ct$. This stage of uninhibited 
quasi-spherical expansion proceeds until the magnetic field pressure 
at the outer edge of the expanding magnetosphere becomes as small as 
the external gas pressure. In order to estimate when this happens, we 
need to evaluate the toroidal magnetic field pressure at $R=R_0(t)$. 
The toroidal field changes with time because of two opposing factors: 
continuing injection of the toroidal magnetic flux, 
$\chi(t)\sim\Psi_0\Omega_* t=\Psi_0 R_0(t)/R_{\rm LC}$,
and the the increasing volume of the cavity. 
The net result is that the toroidal field drops off according to 
\beq
B_\phi[R_0(t),t]\sim {\chi(t)\over{R_0^2(t)}} \sim
{\Psi_0\over{R_{\rm LC}R_0(t)}}\simeq {\Psi_0\over{R_{\rm LC}ct}} \, .
\label{eq-Bphi-free-expansion}
\eeq
Another way to obtain this estimate is to note that the main result 
of this free expansion is the establishment of the stationary isolated-pulsar
magnetosphere inside the radius~$R_0(t)$. The toroidal magnetic field 
in an isolated pulsar magnetosphere scales as $B_\phi(r)\sim 
\Psi_0/R_{\rm LC} r$ (Goldreich \& Julian 1969), which is equivalent 
to the above estimate.

Eventually, the pressure of the toroidal magnetic field drops 
to a level where it becomes equal to the external gas pressure
(there is also a comparable contribution from the electric field). 
This happens at time $t=t_{\rm eq}$, corresponding to the cavity 
radius reaching an equilibrium value
\beq
R_{\rm eq}=ct_{\rm eq} \equiv 
{\Psi_0\over{\sqrt{8\pi P_{\rm ext}}R_{\rm LC}}} \, .
\eeq

After this, the expansion continues, but changes its character:
the lateral expansion slows down, and the expansion becomes mostly 
vertical. Eventually, at $t\gg t_{\rm eq}$, a magnetic tower forms, 
similar to Lynden-Bell's (1996) tower. One important difference is 
that since the radius of the tower is much larger than~$R_{\rm LC}$,
a proper analysis requires relativistic treatment, so Lynden-Bell's 
non-relativistic theory is not directly applicable. In particular, 
we expect the vertical expansion of the tower to be relativistic. 
This can be seen from the following argument. As the tower grows, 
its radius stays roughly constant, of order~$R_{\rm eq}$, whereas 
its height increases linearly with time, with the velocity~$V_{\rm top}$. 
The continuously injected toroidal flux goes into filling the expanding 
volume of the tower with toroidal magnetic field, so that, roughly speaking,
\beq
\chi=\Psi_0 \Omega_* t \sim B_\phi\, V_{\rm top}\, t\, R_{\rm eq}\, .
\eeq
Assuming $B_\phi\sim \sqrt{8\pi P_{\rm ext}}$, we therefore arrive 
at the estimate
\beq
V_{\rm top} \sim c \, .
\eeq
This result can be understood naturally by noting that the problem 
has no mass or density parameter and so there is no characteristic 
velocity scale other than the speed of light~$c$ (scales like 
$\Omega_* R_{\rm eq}$ are even larger than~$c$).

The toroidal magnetic field stays roughly constant during this stage, 
and so the poloidal current flowing through the tower is also constant
and is of order 
\beq
I_{0,\rm eq} \sim B_{\phi,\rm eq} R_{\rm eq}\sim {\Psi_0\over{R_{\rm LC}}}\, ,
\eeq
the same as the poloidal current in the isolated pulsar case.
Correspondingly, the magnetic luminosity (i.e., the spin-down 
power of the pulsar) stays at a constant level of order $P_{\rm isolated}$. 
However, unlike the isolated pulsar case, this luminosity is not 
quasi-spherical, but is channeled predominantly in the vertical
direction.

Provided that the expansion of the tower is sub-magnetosonic,
an approximate relativistic force-free equilibrium is established
inside the tower (at least away from the top lid of the tower). 
As in the fixed-cavity case, the work done by the toroidal field's 
magnetic tension on the equatorial current sheet goes into accelerating 
the equatorial plasma to ultra-relativistic velocities. The relativistic 
mass of this plasma, and hence also the radial centrifugal force grow 
linearly with time, as does the overall magnetic pressure force on the 
outer wall (because of the steadily increasing height of the tower).


\section{Magnetar inside a collapsing star: an outline of the general scenario}
\label{sec-stalled-shock}

\begin{figure}
\plotone{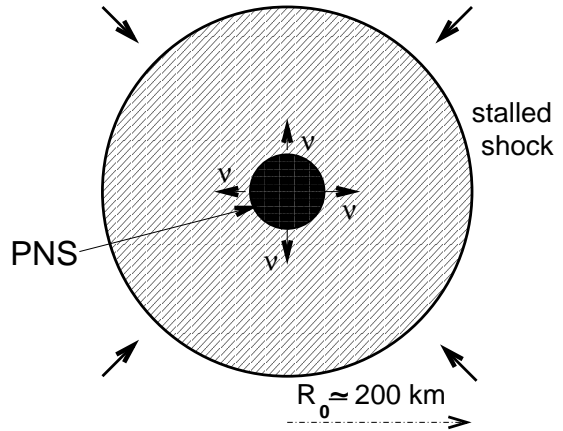}
\figcaption{Stalled shock phase of core-collapse explosion.
\label{fig-stalled-shock}}
\end{figure}

Previous studies of core-collapse supernovae (SNe) have shown that,
when the core of a massive star collapses into a proto-neutron star, 
a bounce shock is launched back into the star (see the reviews by, e.g.,
Bethe \& Wilson~1985; Woosley \& Weaver 1986; Bethe~1990). However, as 
was also shown in these studies, the shock quickly stalls at a radius of 
about 200~km. The explosion then enters a relatively long ($\sim$ 1~sec) 
quasi-stationary phase (see Fig.~\ref{fig-stalled-shock}). During this 
phase accreting material constantly moves through the shock and gets heated 
to very high temperatures. The shock looks stationary in the Eulerian frame 
and the shock jump condition can be viewed as a balance between the ram 
pressure of the infalling material, that tends to quench the shock, and 
the thermal pressure of the post-shocked gas, that is supported mostly by 
the continuous heating due to neutrino deposition in the dense plasma 
behind the shock. Gradually, both the neutrino luminosity and accretion 
rate decline with time. Eventually, one of two things has to happen as 
an outcome of the competition between neutrinos and accretion. 
If neutrinos win, the shock engulfs the entire star and one gets 
a successful SN explosion. If they lose, the shock dies and the PNS 
gains mass beyond the critical mass and collapses into a black hole, 
which then subsequently swallows the rest of the star, without a~SN.

\begin{figure}
\plotone{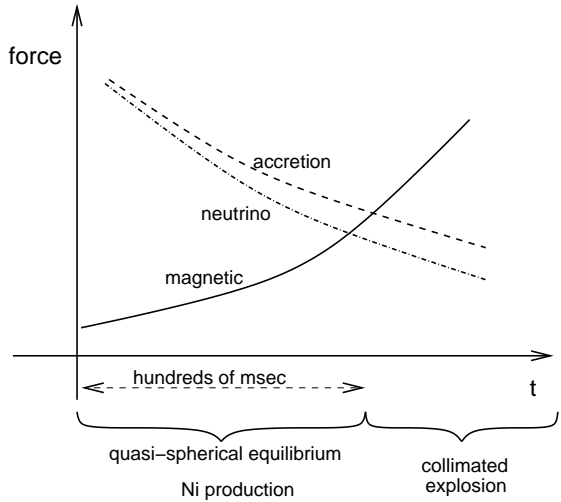}
\figcaption{Schematic time evolution of the main three forces responsible 
for the stalled-shock force balance.
\label{fig-scenario}}
\end{figure}

In our model, we add a third dynamical component --- the magnetic field.
The magnetic force is pushing out, helping the explosion, as is the thermal
pressure of the neutrino-heated gas. These two outward forces are opposed
by the accretion ram pressure. Our main idea is that, generally speaking, 
the two outward forces evolve differently with time, and thus the explosion
may be a two-stage process. In particular, we suggest that the magnetic 
pressure force is unimportant during the stalled-shock phase that lasts
a few hundreds of msec. However, we note that during this time the magnetar
makes several hundred revolutions, resulting in a great amplification of 
the toroidal magnetic flux by the differential rotation. 
[Of course, during this stage the field is not force-free, and the gas 
pressure and inertia are important.] Over time, however, both the neutrino 
energy deposition and the accretion rate decline, whereas the toroidal 
magnetic field grows (see Fig.~\ref{fig-scenario}). 
For example, assuming $R_0=3 R_{\rm LC}= 10 R_*=100$~km, and $B_*=10^{15}$~G, 
we see that the entire cavity is filled with $3\cdot 10^{14}$~G fields after 
about 100~turns (0.1~sec), corresponding to the magnetic pressure of about 
$4\cdot 10^{27}$ erg/cm$^3$. 
This is to be compared with the ram pressure of the infalling 
stellar material that tries to compress the magnetosphere. 
The simplest estimate of the ram pressure at $r=R_0$ is given 
by 
\beq
P_{\rm ram} \sim {{\dot{M}\, v_{\rm ff}}\over{4\pi R_0^2}} \simeq
8 \cdot 10^{27}\, \dot{M}_0\, M_0^{1/2}\, R_{0,7}^{-5/2} {\rm erg\ cm^{-3}}\, ,
\label{eq-P_ram}
\eeq
where $v_{\rm ff}=(2GM/R_0)^{1/2}\simeq 5\cdot 10^9\, M_0^{1/2}\, 
R_{0,7}^{-1/2}\, {\rm cm/sec}$ is the free-fall velocity at radius~$R_0$,
and~$M_0$ and~$\dot{M}_0$ are the mass enclosed within radius~$R_0$ and 
the accretion rate at this radius, expressed in units of $M_{\sun}$
and~$M_{\sun}/{\rm sec}$, respectively.
Thus, after a few hundreds of milliseconds, the magnetic pressure 
overtakes the rapidly decreasing neutrino heating as the main driving
force and re-energizes the stalled shock, leading to a successful 
explosion. This scenario is consistent with the picture presented 
by Akiyama~et~al.\ (2003) who demonstrate numerically the growth of 
the magnetic field on the 200~msec timescale up to about $10^{15}$~G 
in the range of radii up to 100~km. The overall outcome of scenario 
is also similar to that suggested by Bucciantini~et~al. (2006), although, 
because of the winding-up amplification, the magnetic field becomes 
dynamically important much sooner in our model.

To summarize our picture, the ram pressure of the accreting material 
provides a nurturing womb in which the baby magnetic field grows, 
until it is finally strong enough to break out. Neutrino energy 
deposition plays an important role during this gestation period, 
as it provides the support that prevents the magnetosphere from 
being completely squashed and buried by the accreting gas.
Finally, if the above picture is correct and the explosion does 
become magnetically-driven, then the hoop-stress mechanism makes 
it highly collimated, thus satisfying one of the key necessary 
conditions for GRB. Note that this jet is driven by the magnetar-level
(i.e., $\sim10^{15}$~G) field and is therefore stronger and faster than 
the LeBlanc-Wilson (1970) jet that may have been launched a few seconds 
earlier, during the core-collapse process (Wheeler~et~al. 2000).


\section{Discussion}
\label{sec-discussion}


\subsection{Effect of MHD Instabilities}
\label{subsec-stability}

The physical picture presented in this paper, with its smooth coherent
magnetic structure, is, of course, an idealization, necessary for
obtaining a basic physical insight into the system's dynamics and 
for getting the main ideas across in the clearest possible way.
The actual magnetic field is likely to be different from such a simple system
of nested axisymmetric flux surfaces. Instead, it may consist of many 
loops of different sizes and orientations. It may thus have a complex 
substructure on smaller scales, both temporal and spatial. 
This substructure may arise naturally from the beginning, 
especially if the proto-magnetar's magnetic field is produced 
by a turbulent dynamo. On the other hand, it may also result from 
a nonlinear evolution of various MHD instabilities that may develop 
in the system. The effect of MHD instabilities in our highly twisted 
magnetosphere is one of greatest uncertainties in our model.
This section is devoted to the discussion of two such instabilities: 
kink and Rayleigh--Taylor.

({\it i}) 
As the confined magnetosphere is twisted up, it may become prone to 
a non-axisymmetric {\it kink}-like instability. This may happen both 
during the pulsar-in-a-cavity phase and during a later magnetic tower 
phase.

The kink is probably the most dangerous instability in our scenario.
In its nonlinear stage, it may lead to a significant disruption. 
Such a disruption, however, is not necessarily a bad thing: it is 
likely to be only temporary and the tower may be able to reform 
after being disrupted, as is seen in laboratory experiments by 
Lebedev~et~al.\ (2005). The resulting non-steady evolution may 
then provide a plausible mechanism for rapid variability seen in 
gamma-ray bursts. In addition, as a result of such disruption, 
a fraction of the toroidal magnetic field energy may be dissipated 
into thermal energy (Eichler~1993; Begelman~1998). As was shown by 
Drenkhahn \& Spruit (2002; see also Giannios \& Spruit 2005, 2006, 2007), 
this may contribute to the acceleration of the Poynting-flux dominated 
outflow and to powering the prompt gamma-ray emission at later times. 
Also, such kink-driven magnetic dissipation in the magnetosphere may be 
seen as a manifestation of ``coronal activity'' that may modify (harden) 
the emitted neutrino spectrum (Ramirez-Ruiz \& Socrates~2005).

As far as we know, the stability of the pulsar-in-a-cavity has not yet 
been studied. However, several non-relativistic 3D MHD simulations 
(Kato~et~al.\ 2004b; Nakamura~et~al.\ 2007; Ciardi~et~al.\ 2007) have 
recently addressed the kink instability of magnetic towers (although 
not in the GRB context). They seem to indicate that during the first few 
rotation periods, a tower is stabilized by the surrounding high-pressure 
gas, but at later times a large-scale external kink does develop. 
As a result, the tower's overall shape becomes helical. 
This, however, does not immediately lead to the total disruption of 
the tower; even though the configuration is nonaxisymmetric, its main 
morphological features remain similar to those in the axisymmetric 
case (Nakamura~et~al.\ 2007). 
Similar conclusions have been reached by Nakamura \& Meier (2004)
in their 3D-MHD study of Poynting-flux-dominated jets propagating 
through a stratified external medium. These authors found that the 
jet stability strongly depends on the background density and pressure 
profiles along the jet. In particular, a steep external pressure 
gradient forestalls the instability onset. When the instability 
does eventually develop, the resulting helical structures saturate 
and do not develop into full MHD turbulence.
An important theoretical evidence supporting external pressure 
stabilization follows from K\"onigl \& Choudhuri's (1985) analysis 
of a force-free magnetized jet confined by an external pressure. 
They argued that a non-axisymmetric helical equilibrium state becomes 
energetically favorable (conserving the total magnetic helicity in the 
jet) only when the pressure drops below a certain critical value. 
If this happens and the external kink mode does become unstable, 
then this non-axisymmetric equilibrium can be interpreted as 
the end point of the non-linear development of the instability.

In addition to the above non-relativistic studies, a few first steps have
recently been taken towards understanding the stability of relativistic jets, 
in particular, in the framework of relativistic force-free electrodynamics
(Gruzinov~1999; Tomimatsu~et~al.\ 2001). However, to the best of our knowledge,
to date there have been no formal stability studies of relativistic magnetic 
towers or of confined pulsar magnetospheres. Such studies, both analytical 
and numerical, are clearly needed. They may involve a linear perturbation 
analysis or a non-axisymmetric relativistic MHD or force-free simulation. 
They would have to take into account several stabilizing effects. First, 
as Tomimatsu~et~al.\ (2001) found in their linear stability analysis of 
a narrow rotating relativistic force-free jet, rapid field-line rotation 
inhibits kink instability. Second, regarding the stability of a rapidly 
growing magnetic tower, we expect that the tower expansion should quickly 
transition to the relativistic regime, eventually reaching a very large 
$\gamma$-factor. Once this happens, the relativistic time delay may 
effectively stabilize the outflow (see Giannios \& Spruit 2006). 
This is because MHD instabilities grow on the local Alfv\'en-crossing 
time in the fluid frame and hence much slower in the laboratory frame. 
As a result, even if instabilities are excited, they do not have enough 
time to develop before the break-out of the flow from the star.

\begin{figure} [t]
\plotone{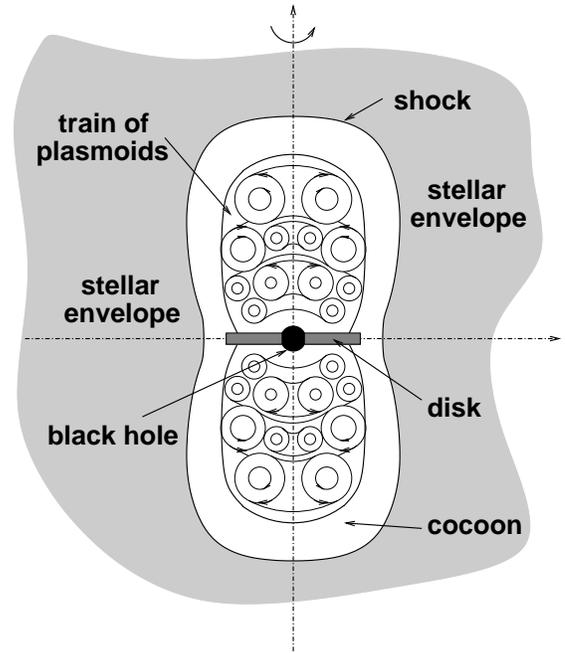}
\figcaption{Magnetic tower may have a substructure represented
by a train of many spheromak-like plasmoids. This situation may 
arise as a result of spatial and temporal intermittency at the 
base of the outflow and/or due to instabilities and reconnection.
\label{fig-train}}
\end{figure}

We would like to point out that, in this problem, we are actually 
interested not so much in the instability onset or its early linear 
development, but rather in its long-term (many rotation periods) 
nonlinear evolution and its overall effect on the magnetosphere. 
Such a long-term behavior is very poorly understood and needs to 
be investigated in the future. Therefore, here we can only provide 
a hypothetical discussion. In its nonlinear stage, the kink instability 
may lead to conversion of some of the toroidal magnetic flux to poloidal 
flux (in our geometry). Some of this new poloidal flux may become detached 
from the star via reconnection (which, in reality, may be strongly inhibited 
deep inside the collapsing star, see Uzdensky \& MacFadyen 2006).
This could lead, in principle, to the break up of the single coherent 
magnetosphere into a number of smaller spheromak-like plasmoids
(Fig.~\ref{fig-train}). 
That is, instead of further twisting up of the entire magnetosphere 
or, at a later stage, lengthening of the magnetic tower, one would 
effectively get continuous injection of new plasmoids into the system. 
Hoop stress still works inside each of them, and so the overall dynamical 
effect may be the same as that of a single magnetosphere, at least 
qualitatively. The resulting multi-component structure of the outflow
may be responsible for the observed intermittency in~GRBs. It is a very 
interesting scenario that should to be considered in future research. 

An important consideration that then needs to be taken into account 
is the conservation of magnetic helicity. Differential rotation leads 
to a continuous injection of helicity into the system (of opposite signs 
in the two hemispheres). The kink instability may convert toroidal flux 
to poloidal but it will not destroy the magnetic helicity accumulated 
in the cavity. Thus, whatever the resulting configuration might be, it 
will have to be consistent with a growing amount of helicity. 
One may in fact imagine a cyclic process involving twisting up the 
magnetosphere for several rotation periods, followed by flux conversion 
due to the kink instability, followed by reconnection and the production 
and detachment of a plasmoid carrying the magnetic helicity (and some of 
the magnetic energy) injected during the given cycle. One can hypothesize 
that if this cyclical process is robust, then over time the bulk of the 
cavity may become filled with spheromak-like plasmoids. The picture then 
would be somewhat similar to that in Figure~\ref{fig-train}, except it
would not have to be axisymmetric. Each of these plasmoids would have 
some net magnetic helicity and magnetic energy and would be in a 
magnetostatic equilibrium configuration, confined laterally by the 
overall pressure of the neighboring plasmoids. It may also contain 
thermal energy produced as a result of reconnection during plasmoid 
creation. Since the magnetic field is closed within each plasmoid, 
each plasmoid is not magnetically connected to the star and thus 
the magnetic field inside of it is not subject to any additional 
twisting. Helicity then stays constant within each plasmoid. As 
the number of such plasmoids grows with time, they occupy larger 
and larger fraction of the cavity volume. Correspondingly, the part 
of the volume that is directly connected to the rotating neutron star 
shrinks with time. As long as some part of the neutron star field lines 
extends beyond the light cylinder, differential rotation continues on 
these field lines, resulting in more twisting and more generation of 
plasmoids. At some point, however, the part of the magnetosphere that 
is directly connected to the neutron star --- the NS magnetosphere proper 
--- will be squeezed by the surrounding plasmoids to such a degree that 
it will be confined entirely inside the light cylinder. If this happens, 
twisting will stop, the neutron star's proper magnetosphere will be 
corotating with the neutron star and magnetic energy extraction will 
cease, at least if the boundary between the proper magnetosphere and 
the surrounding plasmoids is axisymmetric. If this boundary is not 
axisymmetric, then energy and angular momentum will continue to be 
extracted at some level through a process akin to the magnetic propeller 
effect (Illarionov \& Sunyaev 1975).
On the other hand, it is possible that magnetic reconnection will restore 
the link between the magnetosphere proper and the outer plasmoids, even if 
temporarily. If this happens, the situation will become more complicated; 
the spin-down torques will again be modified. In fact, as was pointed out 
to us by the referee of this paper, this resulting situation may become 
somewhat analogous to the case of a pulsar magnetosphere compressed by 
a strong relativistic wind of another pulsar, such as in the case of the 
double pulsar PSR J0737-3039. Such magnetosphere-wind interaction, along 
with the resulting pulsar spin-down torque, was considered by Lyutikov 
(2004) and by Arons~et~al. (2005). As they point out, the torque due to 
the reconnected field lines can become much larger than the usual spin-down 
torque of an isolated pulsar.

It is also interesting to make the following comment. An unbounded 
relativistic force-free outflow driven by a rotating conductor is 
expected to be stable. On the other hand, a closed confined magnetosphere 
with field lines subject to differential rotation, such as our 
pulsar-in-a-cavity problem or a magnetic tower, may be kink-unstable. 
At the same time, as we discussed in Sec.~\ref{subsec-collimation}, 
the outflow is uncollimated in the first case but is collimated in 
the second case. This suggests that there may be a deep connection 
between stability and lack of collimation for axisymmetric relativistic 
force-free flows.

To sum up, the kink, and especially its nonlinear outcome, 
is a serious issue that needs to be addresses in the future. 
Axisymmetric mode (i.e., sausage instability) also needs to 
be investigated.

({\it ii}) 
Another important process that may potentially plague the development of 
coherent magnetic structures inside collapsing stars is the development
of the {\it Rayleigh-Taylor} instability (or, rather, its magnetic 
counter-part, the Kruskal--Schwarzschild instability). This instability 
is expected to affect purely hydrodynamic fireball models as well; 
a strong magnetic field will suppress it somewhat, although probably 
not completely. As Wheeler~et~al. (2000) and Arons (2003) have pointed out, 
the Rayleigh--Taylor instability at the interface between the lightweight 
relativistic fluid (electromagnetic field and hot relativistic plasma) 
and the overlying colder, denser stellar material may cause splitting 
of a coherent magnetic structure into several separate strands interlaced 
with stellar matter. In the Arons (2003) model, the stellar envelope is 
quickly ``shredded'' by the nonlinear Rayleigh-Taylor ``fingers''. 
In effect, these fingers play a role of evacuated channels that allow 
the electromagnetic relativistic plasma energy produced near the central 
engine to escape through the star. Arons further argued that these channels 
suffer only a relatively small amount of mixing with the non-relativistic 
stellar material due to the Kelvin-Helmholtz instability.
In light of this work, we cannot rule out the possibility that our magnetic 
cavity and/or the subsequently-formed magnetic towers may also suffer from
fragmentation into several Rayleigh-Taylor ``fingers''. However, initially 
small-scale fingers quickly merge with each other to form a small number of 
large ones in the nonlinear stage. Therefore, we do not expect strong mixing 
of the baryons from the stellar envelope into the magnetosphere. 
The exact geometry of the outflow may change and a strong time-variability 
may develop, but, overall, we expect the outflow to survive. More research 
is needed in order to assess the implications of this instability for our 
scenario.

Finally,  we would like to reiterate that a proper treatment of these 
problems requires a time-dependent 3D relativistic force-free or full 
(preferably relativistic) MHD analysis and simulations (see~Sec.~\ref
{subsec-numerical}).


\subsection{Nickel Production}
\label{subsec-nickel}

A central issue for the central engine of long-duration GRBs is 
the required production of $^{56}$Ni.  The supernovae that have 
been observed to accompany long duration GRBs (SN-GRBs) are classified 
as Type~Ibc (SNe~Ibc; see, e.g., Soderberg~2006; Kaneko~et~al.~2007). 
Modeling of the optical light curves of SNe Ibc requires the presence 
of radioactive $^{56}$Ni to heat the ejecta after initial post-explosion 
expansion of the star.  The $^{56}$Ni masses inferred from the peak 
optical brightness of SN-GRBs have a broad range, with the brightest, 
e.g., SN1998bw and SN2003dh, requiring several 0.1 M$_{\odot}$.  
On average, however, SN-GRBs are not required to produce more $^{56}$Ni 
than the local population of SNe~Ic (Soderberg 2006).  In fact, as with 
low luminosity~SNe, e.g., the ``tailless'' SnII, some GRB-SNe may produce 
little or no $^{56}$Ni (MacFadyen~2003), as recent observations indicate 
for GRB060505 and GRB060614, two relatively nearby ($z\sim0.1$) long GRBs 
with no detected supernova component (Fynbo~et~al.~2006; Della~Valle et
al.~2006; Gal-Yam~et~al.~2006).

In models of (non-GRB producing) core collapse supernovae, $^{56}$Ni is
produced hydrodynamically in material heated to $T \gtrsim T_{\rm Ni} \sim 
5\times 10^9$~K by the explosion shock launched in the core of the star.  
The amount of $^{56}$Ni produced depends on the mass inside of the expanding
shock when its temperature declines below~$T_{\rm Ni}$. This occurs when its
radius has expanded to 
\beq
R_{\rm Ni}\sim\left(\frac{3E}{4\pi a T_{\rm Ni}^4}\right)^{1/3} \sim
3.7\times 10^8\, E_{51}\, {\rm cm} \, , 
\eeq
where $E=E_{51}\times 10^{51}$~erg is the explosion energy and~$a$ is 
the radiation constant. The mass inside this radius depends on the density
structure of the progenitor star and on how much expansion or contraction 
occurs before the shock reaches a given mass element. In particular, little 
or no $^{56}$Ni is produced by a shock, even if extremely powerful, if it 
is launched into a low density environment.  This may occur if a weak 
initial explosion expands the stellar core so that little mass remains 
within a few $10^8$~cm when the strong shock arrives. 
Production of $\sim 0.1 M_\odot$ of $^{56}$Ni occurs for many pre-supernova 
stars if $\sim 10^{51}$~ergs is deposited isotropically by a (quasi-)spherical 
shock on a timescale of $\sim 1$s so that little pre-expansion of the star 
occurs before the shock arrives.
Some of the brightest supernovae, e.g., SN1998bw, require energies of up to
$\sim 10^{52}$~ergs to make the $\sim 0.5 M_\odot$ inferred from light-curve
modeling.

The requirement of fast ($\lesssim 1$s), isotropic deposition of energy 
for hydrodynamical production of $^{56}$Ni presents a serious challenge 
for models of the SN-GRB central engine.  First, because the GRB engine 
must typically last 10~s or more for relativistic ejecta to escape the star 
and, second, because GRBs are believed to be highly asymmetric explosions.  
The high degree of beaming and long timescale for energy deposition renders 
collapsar jets themselves incapable of producing anywhere near the required 
$^{56}$Ni masses (MacFadyen \& Woosley 1999), since little mass 
($<0.001~M_{\odot}$) can be heated to sufficiently high temperatures.  
Therefore, in the original collapsar model, with a black hole accretion 
disk as the central engine, the $^{56}$Ni is produced in a non-relativistic 
bi-conical wind blown from the disk and constituting a distinct explosion 
component (MacFadyen \& Woosley 1999; MacFadyen~2003).

A fundamental problem for the magnetar model, if it is to produce a GRB and a
supernovae, is the requirement that it produce both an isotropic explosion
for the $^{56}$Ni production and beamed relativistic ejecta.  In our model,
$^{56}$Ni can be produced behind a roughly spherical hydrodynamical shock
driven by the initial quasi-isotropic expansion of the magnetosphere. 
The expansion becomes collimated and the tower formation begins only 
after the stress of the magnetosphere becomes sufficient to balance 
the post-shock pressure. The collimation process of the magnetar wind 
thus involves both a quick isotropic expansion followed by a beamed 
component.  We feel that this modification to the magnetar scenario, 
i.e., the inclusion of the magnetosphere interaction with the exterior 
star, strengthens its viability as a model for the long GRB central
engine.


\subsection{Restarting the Engine}
\label{subsec-restarting}

We note that the same magnetar can power explosions with the degree 
of collimation that depends on the magnitude of the outer bounding
pressure. A quasi-spherical supernova or a highly beamed jet may
result from the same star at different times as the bounding pressure
changes.  In the collapse and explosion of a massive star, the
pressure of stellar gas bounding the central magnetar may have a
complex time history.  The star may initially collimate the embedded
magnetar power into a tightly collimated tower responsible for GRB
emission.  Subsequently, after the star expands and the pressure
bounding the magnetar decreases, the magnetar power will no longer be
strongly beamed and a normal quasi-spherical magnetar outflow will
result.  Later, however, if material not ejected to infinity falls
back and accretes, the magnetar will again be surrounded by a bounding
pressure and its power will be recollimated.  X-ray flares observed
following some GRBs (Burrows~et~al. 2005; Falcone~et~al. 2006; Romano
et~al.~2006) could result from this process (see also, e.g., Proga \&
Zhang 2006; Perna, Armitage \& Zhang 2006).


\subsection{Pre-shaping the cavity}
\label{subsec-preshaping}

In the previous sections, we have shown that toroidal field makes the
expanding plasma self-collimating due to hoop stress, and propose a
spherical cavity with constant wall properties (i.e. no dependence on
polar angle) as the simplest model problem. However, the cavity is
expected in many cases to have lower density near the polar axis at
fixed radius due to various processes acting as the star collapses.
Among these are rotational flattening and the asymmetric stress from
an early magnetized wind. First, in order to produce a millisecond
magnetar, the progenitor star must have been rapidly rotating.  
We therefore expect the collapsed core to be strongly modified by
rotational effects. In particular the material near the rotation axis
experiences no centrifugal barrier inhibiting its accretion, resulting
in a relatively low density in the polar region. A separate effect is
that a weaker, non-relativistic MHD jet may have been launched along 
the axis earlier, during the collapse of the stellar core (LeBlanc \&
Wilson 1970; Wheeler~et~al. 2000). In addition, an initial MHD wind 
from the proto-magnetar may be concentrated to the poles as in 
Bucciantini~et~al. (2006). This will push out the cavity in the polar 
region. The subsequent relativistic wind will then expand into a cavity 
pre-shaped by the previous MHD wind. At a fixed radius, the pressure 
and density of the wall will be decreased at the poles relative to the 
equatorial values. If these effects are extreme, the cavity is significantly 
weakened in the polar direction, and a model problem consisting of a 
``magnetar-in-a-tube'' is of interest.


\subsection{Pulsar Kicks}
\label{subsec-pulsar-kicks}  

Note that in our picture, most of the magnetically-extracted rotational 
energy of the neutron star travels vertically through the two oppositely
directed channels. Correspondingly, a significant amount of linear momentum
is transported up and down from the neutron star and, correspondingly, a 
back-reaction force is exerted on the neutron star from both the top and
the bottom. The two back-reaction forces are oppositely-directed and nearly
cancel each other. However, even a slight imbalance in the force may have
important consequences for the overall momentum impacted to the neutron star
and hence for its terminal velocity. For example, taking the total initial 
rotational energy of the PNS to be $E_{\rm rot}=5\cdot 10^{52}$~erg, the 
momentum transported out in each direction is $P=E_{\rm rot}/2c
\sim 10^{42}$~cgs. Therefore, just a 10\% imbalance would result 
in the terminal velocity of the neutron star of order of $v_{\rm term}
\simeq 0.1 P/M_{\rm NS}\sim 300$~km/sec.


\subsection {Prospects for Numerical Simulations}
\label{subsec-numerical}

In order to gain a solid physical understanding of the fundamental
physical processes controlling the interaction of a magnetar with its
birth environment, we suggest a sequence of numerical investigations
employing a range of well-tested plasma descriptions.  Of particular
usefulness are limiting cases which allow for simplified analysis
making the key physics more transparent. Simulations should cover
regions of parameter space where limiting cases overlap with more
complete plasma descriptions.  For example, force-free (degenerate)
electrodynamics (FFDE) is a useful tool for studying highly magnetized
plasma for which pressure and inertia are negligibly small. In this 
extreme case, the cavity wall would have to be represented by a rigid 
perfectly conducting outer boundary condition. While this case may not 
be of general relevance for the realistic physical environment, some 
basic aspects of a bounded rotating magnetosphere may be understood 
using this description. In addition, the FFDE description has the 
advantage of requiring fewer parameters to specify the initial and 
boundary conditions for a given model problem.  The results of the
time integration can then be more easily understood with a minimum 
of complicating factors. Time-dependent force-free codes have already 
been used successfully in the recent years to study pulsar magnetospheres 
(e.g., Komissarov~2006; McKinney~2006b; Spitkovsky~2006).

It is possible that the full magnetar-in-a-star problem can be
successfully investigated by a hybrid simulation that would employ 
a relativistic force-free code inside the cavity and a relativistic
hydrodynamic simulation outside (e.g., R.~D. Blandford~2005, private
communication, McKinney~2006a).

The next step would be to treat the plasma in the fully relativistic
MHD regime.  There are several relativistic MHD codes in existence
that have reached the required level of maturity (Koide, Shibata \&
Kudoh 1999; Gammie, McKinney \& T\'oth 2003; Del~Zanna, Bucciantini 
\& Londrillo~2003; De~Villiers, Hawley \& Krolik 2003; Fragile~2005;
Komissarov~2005; Nishikawa~et~al.\ 2005). Of interest would be a set
of simulations with a range of plasma~$\beta$. The low-$\beta$
simulations should match on to the FFDE case, at least qualitatively.
Once these simulations are analyzed and the basic physical processes
elucidated, $\beta$ can be gradually increased enabling an
understanding of how plasma inertia and pressure affect the dynamics
of the magnetosphere expansion and collimation.

The basic process of tower formation and collimation can initially be
explored with two-dimensional axisymmetric simulations.  However, to
investigate tower stability to non-axisymmetric disruptions, fully
three-dimensional simulations are necessary.

Finally, note that the general processes we describe here are of interest 
for many astrophysical systems including non-relativistic central objects
(e.g., planetary nebulae, see Blackman~et~al.\ 2001, Matt~et~al.\ 2001). 
For this reason, non-relativistic MHD simulations of this problem are 
of interest in themselves, as well as a first step toward fully 
relativistic MHD.
Recent non-relativistic MHD simulations indicate that the magnetic tower 
mechanism can operate successfully in a variety of astrophysical environments 
(e.g., Romanova~et~al.\ 2004; Kato~et~al.\ 2004ab; Nakamura~et~al. 2006, 2007).


\section{Conclusions}
\label{sec-conclusions}

In this paper we have investigated the millisecond-magnetar scenario 
for the central engine of Gamma-Ray Bursts and core-collapse Supernovae.
We have focused on the interaction between the rapidly-rotating magnetar
magnetosphere and the surrounding infalling stellar envelope. We have 
argued that the stellar material provides a confining (ram) pressure
that has a strong effect on both the size and the shape of the magnetosphere.
In particular, it can channel the highly-magnetized outflow originating
from the proto-neutron star into two collimated magnetic towers.

More specifically, we suggest that the stalled bounce shock --- a 
common feature in models of core-collapse supernovae --- effectively 
plays a role of a cavity that confines the magnetosphere. 
The cavity's radius, determined by the balance between the pressure 
of the hot neutrino-heated gas and the ram pressure of the infalling 
material, stays quasi-stationary at $R_0\simeq 200$~km during the first 
few hundreds of milliseconds after the bounce.
To get a feeling for what happens to the magnetar magnetosphere during 
this stage, we introduce a simplified fundamental-physics problem that 
we call the {\it Pulsar-in-a-Cavity} problem. A large part of our paper 
(\S~\ref{sec-pulsar-cavity}) is devoted to investigating this problem.
We show that since the radius of the cavity is larger than the pulsar 
light-cylinder radius, the magnetic field inside the cavity continuously 
winds up. Correspondingly, both the toroidal field strength and the 
magnetic spin-down luminosity of the pulsar increase roughly linearly 
with time. The magnetic energy in the cavity then grows quadratically 
with time. We then demonstrate that in the context of a millisecond 
magnetar inside a collapsing star the magnetic field becomes dynamically 
important a fraction of a second after the bounce. This leads to a 
subsequent revival of the stalled shock and may result in a successful 
magnetically-driven explosion. As long as the expansion of the cavity 
is non-relativistic, the toroidal magnetic field inside it remains larger 
than the poloidal magnetic and electric fields. As a result, the hoop-stress 
collimates the Poynting-flux-dominated outflow into two vertical channels 
that are similar to Lynden-Bell's (1996) magnetic towers (see Uzdensky \&
MacFadyen~2006).

Finally, we discuss the implications of model for several 
observationally-motivated questions relevant to GRBs and 
core-collapse supernovae, such as $^{56}$Ni production, 
late-time X-ray flares, and pulsar kicks. We also outline 
a set of numerical studies that we feel need to be done.


\acknowledgments

We are grateful to A.~Beloborodov, E.~Blackman, J.~Goodman, A.~K\"onigl,
R.~Kulsrud, M.~Lyutikov, J.~McKinney, J.~Ostriker, C.~Thompson, and to
the anonymous referee for encouraging conversations and critical remarks.

AIM acknowledges support from the Keck Fellowship at the Institute 
for Advanced Study. DAU's research has been supported by the National 
Science Foundation under Grant~PHY-0215581 (PFC: Center for Magnetic 
Self-Organization in Laboratory and Astrophysical Plasmas).


\section*{REFERENCES}
\parindent 0 pt

Akiyama, Wheeler, Meier, \& Lichtenstadt\ 2003, ApJ, 584, 954

Aloy, M.~A., M\"uller, E., Ibáñez, J. M., Martí, J. M. \& MacFadyen, A.,
2000, ApJ, 531, 119

Ardeljan, N.~V., Bisnovatyi-Kogan, G.~S., \& Moiseenko, S.~G.\ 2005, 
MNRAS, 359, 333

Arons, J.\ 2003, ApJ, 589, 871

Arons, J., Backer, D.~C., Spitkovsky, A., \& Kaspi V.~M.\ 2005, 
in ASP Conf.\ Ser.~328, Binary Radio Pulsars, ed.\ F.~A.~Rasio 
\& I.~H.~Stairs (San Francisco: ASP), 95 

Begelman, M.~C. 1998, ApJ, 493, 291

Bethe, H.~A., \& Wilson, J.~R.\ 1985, ApJ, 295, 14 

Bethe, H.~A.\ 1990, Rev. Mod. Phys., 62, 801 

Blackman, E.~G., Frank, A., \& Welch, C.\ 2001, ApJ, 546, 288

Blandford, R.~D., \& Znajek, R.~L.\ 1977, MNRAS, 179, 433

Bucciantini, N., Thompson, T.~A., Arons, J., Quataert, E., 
\& Del~Zanna, L.\ 2006, MNRAS, 368, 1717 

Burrows, D.~N. et~al.\ 2005, Science, 309, 1833

Burrows, A., Dessart, L., Livne, E., Ott, C.~D., \& Murphy, J.\ 2007; 
accepted to ApJ; preprint (astro-ph/0702539)

Ciardi, A., Lebedev, S.~V., Frank, A., et~al.\ 2007, 
Phys. Plasmas, 14, 056501

Contopoulos, I., Kazanas, D., \& Fendt, C.\ 1999, ApJ 511, 351

Della Valle, M., Chincarini, G., Panagia, N., et~al.\ 2006, Nature, 444, 1050 

Del~Zanna, L., Bucciantini, N., \& Londrillo, P.\ 2003, A\&A, 400, 397

De Villiers, J.-P., Hawley, J.~F., \& Krolik, J.~H.\ 2003, ApJ, 599, 1238

Drenkhahn, G. \& Spruit, H.\ 2002, A\&A, 391, 1141

Duncan, R.~C., \& Thompson, C.\ 1992, ApJ, 392, L9

Eichler, D. 1993, ApJ, 419, 111

Falcone, A.~D. et al.\ 2006, ApJ, 641, 1010

Fragile, P.~C.\ 2005; preprint (astro-ph/0503305)

Fynbo, J.~P.~U.~, Watson, D., Th\"one, C.~C., et~al.\ 2006, Nature, 444, 1047

Gal-Yam, A., Fox, D.~B., Price, P.~A., et~al.\ 2006, Nature, 444, 1053

Gammie, C.~F., McKinney, J.~C., \& T\'oth, G.\ 2003, ApJ, 589, 444

Giannios, D. \& Spruit, H.\ 2005, A\&A, 430, 1

Giannios, D. \& Spruit, H.\ 2006, A\&A, 450, 887

Giannios, D. \& Spruit, H.\ 2007, A\&A, 469, 1

Goldreich, P. \& Julian, W.~H.\ 1969, ApJ, 157, 869 

Goldreich, P., Pacini, F., \& Rees, M.~J.\ 1971, 
Comments on Astrophysics and Space Physics, 3, 185

Gruzinov, A.\ 1999; preprint (astro-ph/9908101) 

Illarionov, A.~F. \& Sunyaev, R.~A.\ 1975, A\&A, 39, 185

Kaneko, Y., Ramirez-Ruiz, E., Granot, J., et~al.\ 2007, ApJ, 654, 385

Kardashev, N.~S.\ 1970, Sov. Astron., 14, 375

Kato, Y., Hayashi, M.~R., \& Matsumoto, R.\ 2004a, ApJ, 600, 338
	
Kato, Y., Mineshige, S., \& Shibata, K.\ 2004b, ApJ, 605, 307

Koide, S., Shibata, K., \& Kudoh, T.\ 1999, ApJ, 522, 727

Komissarov, S.~S.\ 2005, MNRAS, 359, 801

Komissarov, S.\ 2006, MNRAS, 367, 19

K{\"o}nigl, A. \& Choudhuri, A.~R.\ 1985, ApJ, 289, 173

LeBlanc, J.~M. \& Wilson, J.~R.\ 1970, ApJ, 161, 541

Lynden-Bell, D.\ 1996, MNRAS, 279, 389

Lyutikov, M. \& Blandford, R.\ 2002, Proceeding of the 
Workshop on {\it Beaming and Jets in Gamma Ray Bursts (NBSI)},
ed.~R.~Ouyed, Copenhagen, Aug.~2002; preprint (astro-ph/0210671)

Lyutikov, M. \& Blandford, R.\ 2003; preprint (astro-ph/0312347)

Lyutikov, M.\ 2004, MNRAS, 353, 1095 

Lyutikov, M.\ 2006, New J.\ Phys., 8, 119  

MacFadyen, A.~I. \& Woosley, S.~E.\ 1999, ApJ, 524, 262

MacFadyen, A.~I., Woosley, S.~E., \& Heger, A.\ 2001, ApJ, 550, 410 

MacFadyen, A.~I.\ 2003, ``Gamma-ray Burst and Afterflow Astronomy: 2001'' AIP
Conference Proceedings, 662, 202

Matt, S., Frank, A., \& Blackman, E.~G.\ 2006, ApJ, 647, L45

McKinney, J.~C.\ 2006a, MNRAS, 367, 1797

McKinney, J.~C.\ 2006b, MNRAS, 368, L30

Metzger, B.~D., Thompson, T.~A., \& Quataert, E.\ 2007, ApJ, 659, 561

Michel, F.~C.\ 1973, ApJ, 180, L133 

Nakamura, T.\ 1998, Progress of Theoretical Physics, 100, 921

Nakamura, M., Li, H., \& Li, S.\ 2006, ApJ, 652, 1059

Nakamura, M., Li, H., \& Li, S.\ 2007, ApJ, 656, 721

Nakamura, M. \& Meier, D.~L.\ 2004, ApJ, 617, 123

Nishikawa, K.-I., Richardson, G., Koide, S., Shibata, K., Kudoh, T., 
Hardee, P., \& Fishman, G.~J.\ 2005, ApJ, 625, 60

Ostriker, J.~P. \& Gunn, J.~E.\ 1971, ApJ, 164, L95

Paczynski, B.\ 1998, ApJ, 494, L45

Perna, R., Armitage, P. \& Zhang B., 2006, ApJ 636, L29

Proga, D. \& Zhang, B., 2006, MNRAS, 370, 61

Ramirez-Ruiz, E. \& Socrates, A.\ 2005; preprint (astro-ph/0504257)

Romano, P. et al., 2006, A\&A, 450, 59

Romanova, M.~M., Ustyugova, G.~V., Koldoba, A.~V., \& Lovelace, R.~V.~E.\
2004, ApJ, 616, L151

Ruderman, M.~A., Tao, L., \& Kluzniak, W.\ 2000, ApJ, 542, 243 

Soderberg, A., 2006, ``Gamma-Ray Bursts in the Swift Era, Sixteenth
Maryland Astrophysics Conference'' Eds. Holt, S.S., Gehrels, N. \&
Nousek, J. A., AIP Conference Proceedings, 838, 380.

Spitkovsky, A.\ 2006, ApJ, 648, L51

Spruit, H.\ 1999, A\&A, 341, L1

Thompson, C. \& Duncan, R.~C.\ 1993, ApJ, 408, 194

Thompson, C.\ 1994, MNRAS, 270, 480 

Thompson, T.~A., Chang, P., \& Quataert, E.\ 2004, ApJ, 611, 380 

Tomimatsu, A., Matsuoka, T., \& Takahashi, M.\ 2001, Phys.~Rev.~D., 64, 123003

Usov, V.~V.\ 1992, Nature, 357, 472 
 
Uzdensky, D.~A. \& MacFadyen, A.~I.\ 2006, ApJ, 647, 1192

Wheeler, J.~C., Yi, I., H{\"o}flich, P., \& Wang, L.\ 2000, ApJ, 537, 810

Wheeler, J.~C., Meier, D.~L., \& Wilson, J.~R.\ 2002, ApJ, 568, 807

Woosley, S.~E. \& Weaver, T.~A.\ 1986, ARA\&A, 24, 205

Woosley, S.~E.\ 1993, ApJ, 405, 273

Yi, I. \& Blackman, E.~G.\ 1998, ApJ, 494, L163
	
Zhang, W., Woosley, S.~E., \& MacFadyen, A.~I.\ 2003, ApJ, 586, 356


\end{document}